\documentclass[journal]{IEEEtran}
\ifCLASSINFOpdf
\else
\fi
\usepackage{tikz,pgfplots}
\usetikzlibrary{trees}
\usetikzlibrary{shapes,arrows,calc}

\usepackage{cite}
\usepackage{amsmath,amssymb,amsfonts}
\usepackage{graphicx}
\usepackage{textcomp}

\usepackage{float}
\usepackage{subfig}

\usepackage[linesnumbered,ruled,vlined]{algorithm2e}
\usepackage{algpseudocode}
\usepackage{mathtools}
\usepackage[colorlinks=true,urlcolor=blue,linkcolor=black,citecolor=blue]{hyperref}
\usepackage{glossaries}
\usepackage{multirow}
\usepackage{multicol}

\newacronym{HEVC}{HEVC}{high efficiency video coding}
\newacronym{AVC}{AVC}{advanced video coding}
\newacronym{SHVC}{SHVC}{scalable high efficiency video coding}
\newacronym{SVC}{SVC}{scalable video coding}
\newacronym{CTC}{CTC}{common test conditions}
\newacronym{BL}{BL}{base layer}
\newacronym{EL}{EL}{enhancement layers}
\newacronym{JCT-VC}{JCT-VC}{joint collaborative team on video coding}
\newacronym{MPEG}{MPEG}{motion picture experts group}
\newacronym{ATSC}{ATSC}{advanced television systems committee}
\newacronym{SDR}{SDR}{standard dynamic range}
\newacronym{HDR}{HDR}{high dynamic range}
\newacronym{MVs}{MVs}{motion vectors}
\newacronym{SIMD}{SIMD}{single instruction multiple data}
\newacronym{CU}{CU}{coding unit}
\newacronym{CTU}{CTU}{coding tree unit}
\newacronym{DCT}{DCT}{discrete cosine transform}
\newacronym{DST}{DST}{discrete sine transform}
\newacronym{AI}{AI}{all intra}
\newacronym{RA}{RA}{random access}
\newacronym{LDP}{LD. P}{low delay P}
\newacronym{QP}{QP}{quantization parameter}
\newacronym{BD-BR}{BD-BR}{Bj\o ntegaard delta bit rate}

\newacronym{SHM}{SHM}{SHVC reference software model}
\newacronym{JSVM}{JSVM}{joint scalable video model}
\newacronym{SSE}{SSE}{streaming SIMD extensions}
\newacronym{fps}{fps}{frames per second}
\newacronym{GOP}{GOP}{group of pictures}
\newacronym{MV}{MV}{motion vector}
\newacronym{POC}{POC}{picture order count}
\newacronym{simulcast}{simulcast}{simultaneous broadcast}
\newacronym{CABAC}{CABAC}{context-adaptive binary arithmetic coding}
\newacronym{HFR}{HFR}{high frame rate}
\newacronym{JVET}{JVET}{joint video experts team}
\newacronym{ITU}{ITU}{international telecommunication union} 
\newacronym{JEM}{JEM}{joint exploration model}
\newacronym{IPM}{IPM}{intra prediction mode}
\newacronym{FL}{FL}{fixed length}
\newacronym{MPM}{MPM}{most probable modes}
\newacronym{PU}{PU}{prediction unit}
\newacronym{QTBT}{QTBT}{quad-tree plus binary-tree}
\newacronym{RD}{RD}{rate distortion}
\newacronym{HM}{HM}{HEVC reference software model}
\newacronym{VL}{VL}{variable length}
\newacronym{VVC}{VVC}{versatile video coding}

\newacronym{SPS}{SPS}{sequence parameter set}

\begin{document}
%
\title{Efficient Predictive Coding of Intra Prediction Modes}
%
%
%

\author{Kevin Reuz\'e,
        Wassim~Hamidouche,
        Pierrick Philippe, and~Olivier~D\'eforges
\thanks{Kevin Reuz\'e and Pierrick Philippe were with Orange Labs, 4 Rue du Clos Courtel, 35510 Cesson-S\'evign\'e.} 
\thanks{Wassim Hamidouche and Olivier D\'eforges were with Univ. Rennes, INSA Rennes, CNRS, IETR - UMR 6164, Rennes, France.}
}

%
%

\markboth{}%
{Shell \MakeLowercase{\textit{et al.}}: Bare Demo of IEEEtran.cls for IEEE Journals}
%



\maketitle

\begin{abstract}
The \gls{HEVC} standard and the \gls{JEM} codec incorporate 35 and 67 \glspl{IPM} respectively, which are essential for efficient compression of Intra coded blocks. These \glspl{IPM} are transmitted to the decoder through a coding scheme. In our paper, we present an innovative approach to construct a dedicated coding scheme for \gls{IPM} based on contextual information. This approach comprises three key steps: prediction, clustering, and coding, each of which has been enhanced by introducing new elements, namely, \textit{labels} for prediction, \textit{tests} for clustering, and \textit{codes} for coding. In this context, we have proposed a method that utilizes a genetic algorithm to minimize the rate cost, aiming to derive the most efficient coding scheme while leveraging the available \textit{labels}, \textit{tests}, and \textit{codes}. The resulting coding scheme, expressed as a binary tree, achieves the highest coding efficiency for a given level of complexity. In our experimental evaluation under the \gls{HEVC} standard, we observed significant bitrate gains while maintaining coding efficiency under the \gls{JEM} codec. These results demonstrate the potential of our approach to improve compression efficiency, particularly under the \gls{HEVC} standard, while preserving the coding efficiency of the \gls{JEM} codec.
\end{abstract}

\begin{IEEEkeywords}
HEVC, Intra Prediction Mode (IPM), Predictive coding, Versatile Video Coding (VVC).
\end{IEEEkeywords}

%
\IEEEpeerreviewmaketitle

\glsresetall

\section{Introduction} 
 The latest video coding standard, \gls{HEVC}\cite{sullivan_overview_2012}, which was finalized in 2013, has achieved a remarkable bitrate reduction of 50\% compared to the previous \gls{AVC} standard~\cite{wiegand_overview_2003}, while maintaining the same subjective quality~\cite{ohm_comparison_2012,tan_video_2016}. In pursuit of even greater coding efficiency, the \gls{JVET}, established by \gls{MPEG}ISO/IEC and \gls{ITU}, has explored new coding tools to develop the \gls{VVC} standard. These proposed coding tools~\cite{8281012, 7921897} have been incorporated into the \gls{JEM} software, resulting in coding gains of up to 35\%~\cite{JVET-B0044,JVET-B0022, 8296832} when compared to \gls{HM}~\cite{SHM_reference_2017}. However, it's important to note that this coding gain comes at the expense of significantly increased coding complexity, estimated to be 10 times that of \gls{HEVC}. 

This paper focuses on the signaling of \gls{IPM} in both \gls{HEVC} and \gls{JEM} bitstreams. These hybrid video coding schemes employ Intra and Inter predictions to eliminate spatial and temporal correlations in the video sequence, respectively. In recent video coding standards, Intra prediction is performed at the block level and is applied to both Intra and Inter coded slices. The number of \glspl{IPM} has increased from 14 modes in \gls{AVC}~\cite{10.1007/978-1-4020-8737-0_3} to 35 and 67 in \gls{HEVC} and \gls{JEM}, respectively. These modes cover a wider range of directions towards the reference pixels to enhance the accuracy and efficiency of Intra prediction (See Fig.~\ref{fig:intra_mode_jem}).

However, increasing the number of \glspl{IPM} results in higher coding complexity and an increased number of bits required to encode the selected \gls{IPM} at the block level. Several algorithms have been proposed~\cite{6662471,mercat_energy_2017, 8412615} to reduce Intra prediction complexity by testing a restricted set of \glspl{IPM} with only a slight coding loss. Other complexity reduction techniques focus on predicting block partitioning~\cite{8123876, 8305020}, meaning that Intra prediction modes are tested by the encoder only for certain block configurations. Additionally, signaling more modes would require a higher average number of bits for transmission. To maintain the rate-distortion performance, these modes must be efficiently encoded in the bitstream. In \gls{HEVC} and \gls{JEM}, \glspl{IPM} are efficiently encoded by leveraging contextual information, including the \glspl{IPM} of neighboring blocks that are already available at the decoder side. Based on this contextual information, these solutions construct a list of \glspl{MPM} that are encoded with shorter codes, while the remaining modes are encoded with longer codes. These bits are further compressed using entropy encoding to reduce statistical redundancies.

In this context, authors in~\cite{8351605} have proposed a new ordering of the \textit{labels} used to construct the \glspl{MPM} list in the \gls{JEM} software, based on an offline training process. They defined a total of 4019 possible orderings according to available contextual information, block shape, and depth. This approach resulted in a coding gain of 0.12\% at the expense of increased memory usage. However, the training and testing processes were performed on the same video dataset defined in the \gls{JEM} \gls{CTC}. Further, authors in~\cite{9115218} introduced two \gls{MPM} lists (short-term and long-term lists) for more efficient coding of \glspl{IPM} in the \gls{VVC} standard. The long-term list captures high correlations with non-adjacent \glspl{CU} within the \gls{CTU}, primarily identified in screen content videos. The merge of local and global lists relies on a conditional random field to uniformly model the priority of intra modes. The long-term list is populated using a frequency table of \glspl{IPM}, which is initialized with predefined tables signaled at the \gls{SPS} and then updated with selected modes in the \gls{CTU}. This approach achieved coding gains of 3.75\% for screen content videos and 0.1\% for natural scene video sequences. Nevertheless, it's worth noting that the coding schemes for \glspl{IPM} used in \gls{HEVC} and \gls{JEM} are derived empirically based on experimental results. To the best of our knowledge, there is no explicit solution in the literature that derives an optimal signaling scheme for encoding \glspl{IPM} in video coding standards.

The main contributions of this paper can be summarized as follows:
\begin{itemize}
\item We model the construction of efficient signaling schemes for discrete predictive coding based on contextual information.
\item We utilize this model to create efficient signaling schemes for \glspl{IPM} in both reference software implantation of \gls{HEVC} and \gls{JEM}.
\item Our approach results in a rate-distortion improvement for reference software implantation of \gls{HEVC} under similar complexity conditions at both the encoder and decoder.
\end{itemize}

The remainder of this paper is organized as follows. Section~\ref{Related_works} provides background information on Intra prediction and outlines the motivations behind this work. In Section~\ref{model}, we present our proposed model for optimizing code selection using contextual information. Section~\ref{HMJEM} applies the proposed model to \gls{HEVC} and \gls{JEM} to construct efficient \gls{IPM} signaling schemes. The performance of these schemes is assessed in terms of rate-distortion and complexity in Section~\ref{res}. Section~\ref{sec:dynamiclist} extends the proposed approach to adaptive dynamic lists for more efficient \gls{IPM} coding in the \gls{JEM} reference software. Finally, Section~\ref{concl} concludes the paper.

\section{Background and motivations} 
\label{Related_works}
\subsection{Intra prediction in \gls{HEVC} and \gls{JEM}}
The \gls{HEVC} frame is first partitioned into square blocks called \gls{CTU} of fixed size $N \times N$, with $N \in \{16, \, 32, \, 64\}$~\cite{6324413}. Each \gls{CTU} is partitioned in Quad-Tree to \glspl{CU} which are then split into \glspl{PU}. The \gls{PU} represents the base unit for both Intra and Inter predictions. In the case of Intra prediction, the \glspl{PU} are square of sizes $N \times N$ and $2N \times 2N$ with the associated \gls{CU} of size $2N \times 2N$, $N\in \{ 4, 8, 16 \}$. \gls{HEVC} defines 35 \glspl{IPM} including 1 planar mode, 1 DC mode and 33 directional modes to predict the \gls{PU} pixels from the neighbour decoded pixels~\cite{6317153}. The \glspl{IPM} are signalled in the bitstream to the decoder at the \gls{CU} level and then all \glspl{PU} associated to this \gls{CU} share the same \gls{IPM}~\cite{7479465}. 


In the \gls{JEM} codec, the Quad-Tree partitioning used in \gls{HEVC} is replaced by the \gls{QTBT}~\cite{7921897} partitioning. The \gls{QTBT} partitioning first performs a Quad-Tree partitioning of the \gls{CTU} into \glspl{CU} which are further partitioned into blocks by Binary-Tree. This later enables horizontal and vertical symmetrical partitioning, which results in more flexible patterns for splitting the blocks leading to better coding efficiency~\cite{8123876}. The blocks corresponding to the splitting of the Binary-Tree are the basis units to perform both Inter and Intra predictions. The Intra prediction in the \gls{JEM} is performed on blocks of different shapes and sizes. There are in total  67 \glspl{IPM} including the planar, DC and 65 directional modes. 


\subsection{\gls{IPM} coding} 


\begin{figure}[t]
	\centering
	\subfloat[Intra Prediction modes]{\includegraphics[width= 0.65\linewidth]{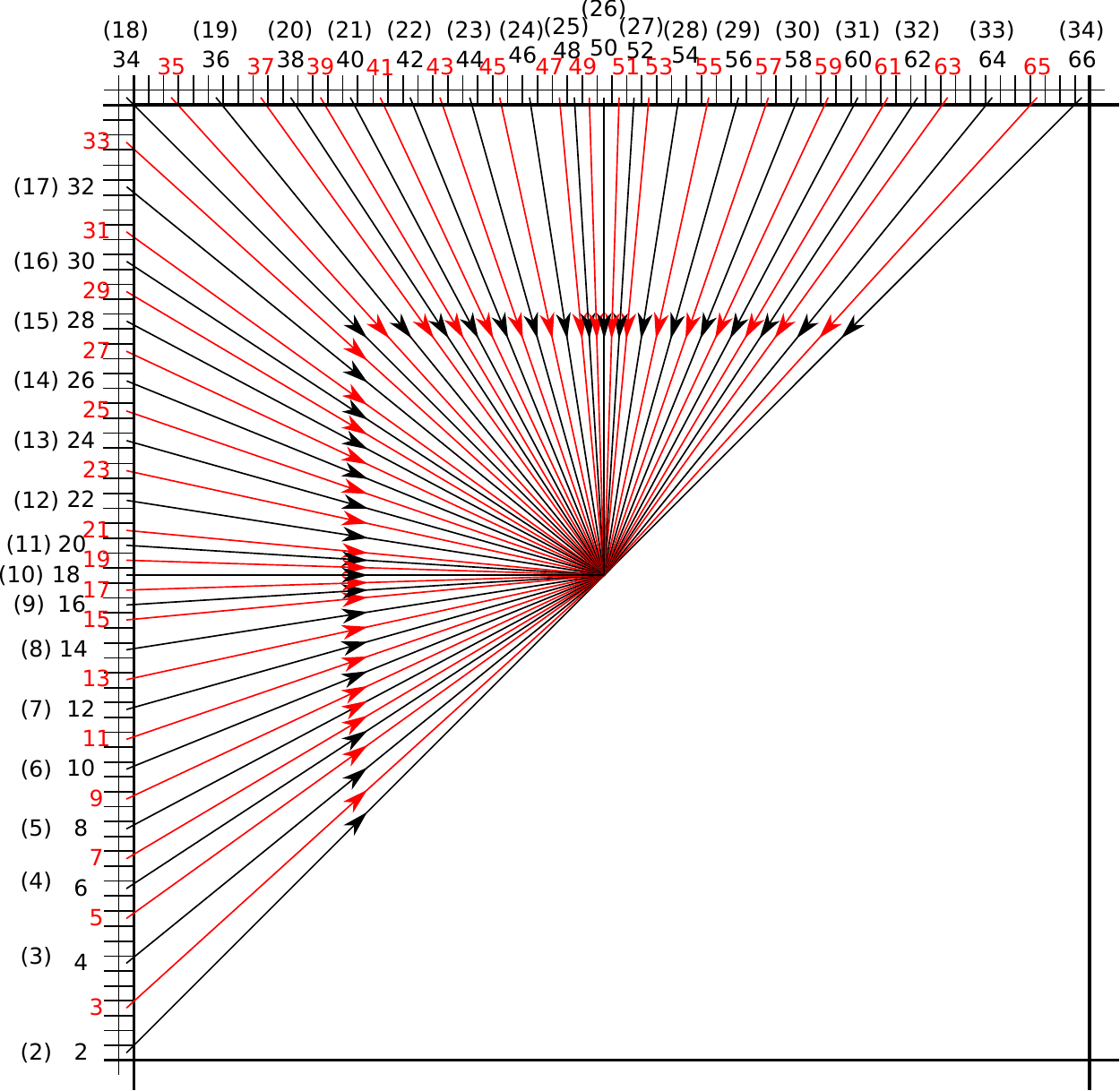}\label{fig:intra_mode_angles_jem}}
	\subfloat[Neighbour blocks]{\includegraphics[width= 0.33\linewidth]{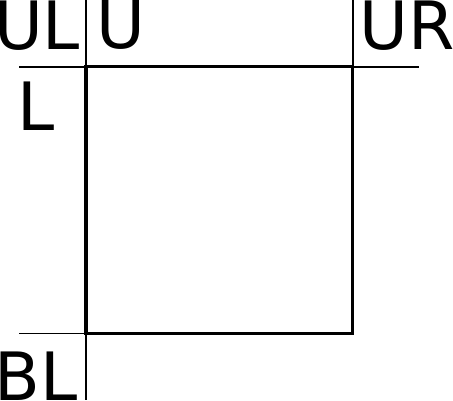}\label{fig:l_u_bl_ur_ul_positions}} 
	\caption{(a) Angles of the 65 directional \glspl{IPM} as defined in \gls{JEM}. The angles in black represent the 33 directional modes available in \gls{HEVC}, with their numbering from \gls{HEVC} in parenthesis, and in red the new modes added in the \gls{JEM}. (b) Positions of blocks coded with \gls{IPM} $L$, $U$, $BL$, $UR$, $UL$}
	\label{fig:intra_mode_jem}
\end{figure}

\gls{HEVC} uses predictive coding to encode the 35 possible \glspl{IPM}. The \glspl{IPM} of left block ($L$) and upper block ($U$) with other numerical modes including 0 (planar), 1 (DC) and 26 (vertical) are used to predict the \gls{IPM} of the current block and construct its \glspl{MPM} list of three modes. The first \gls{MPM} (\gls{MPM}0) is encoded with a code of two bits length, the second and third \glspl{MPM} are signalled by codes of three bits while the remaining 32 \glspl{IPM} are encoded with codes of 6 bits. Table~{\ref{tab:hevc_IPM_codes}} gives the codes of the \gls{IPM}s as defined in the \gls{HEVC} standard. The first bit of each code, illustrated in red color in Table~{\ref{tab:hevc_IPM_codes}}, is encoded with the \gls{CABAC} context~\cite{6317157}. The \gls{IPM} coding scheme in \gls{HEVC} is noted in the rest of this paper as $2+3+3+(6 \times  32)$. This notation gives the number of bits of the three \glspl{MPM} followed by the number of bits used to code the remaining modes times the number of no \glspl{MPM}.

The \gls{JEM} also uses the contextual information to predict the current \gls{IPM} and construct the \glspl{MPM} list. The construction of this \glspl{MPM} list is based on four contexts. As illustrated in Fig.~{\ref{fig:l_u_bl_ur_ul_positions}} these contexts are the \gls{IPM} used by the blocks left ($L$), upper ($U$), below left ($BL$), upper right ($UR$) and upper left ($UL$). 
The \gls{JEM} defines four specific lists to efficiently encode the \glspl{IPM} called: \glspl{MPM} list of 6 modes, \textit{Preferred} list of 16 modes,  \textit{other part 1} list of 19 modes and \textit{other part 2} list that includes 26 modes. The construction of the \glspl{MPM} list is filled by including, in this order,  modes $L$, $U$, Planar, DC, $BL$, $UR$,  $UL$, derived modes ($-1$, $+1$ of the already selected \glspl{MPM}) and numerical \glspl{IPM} including $50$ (vertical), $18$ (horizontal), $34$ (diagonal) and $2$ (diagonal). The directions of the 65 angular modes in \gls{JEM} are shown in Fig.~{\ref{fig:intra_mode_angles_jem}}. These modes are included in the \glspl{MPM} list until this list is filled with 6 different modes. The \textit{Preferred} list contains modes of indexes from $0$ to $60$ modulo $4$ (i.e. indexes $\in \{0, \, 4, \, 8, ... , \, 56 , \, 60 \}$) and not selected as \glspl{MPM}. The modes of indexes from $0$ to $18$ not present in the \glspl{MPM} and \textit{Preferred} lists are used to fill the \textit{other part 1} list. Finally, the modes of indexes from $19$ to $45$ and not present in the three previous lists are used to fill the \textit{other part 2} list.   
The 6 \glspl{MPM} are encoded by 2, 3, 4, 5, 6, and 6 bits, respectively. The 16 \glspl{IPM} in the \textit{Preferred} list are coded on 6 bits. The \glspl{IPM} of the \textit{other part 1} and \textit{other part 2} lists are coded on 7 and 8 bits, respectively. The coding of \glspl{IPM} in the \gls{JEM} is summarized in Table~{\ref{tab:JEM_IPM_codes}}. The first bits of each mode illustrated in red color in Table~{\ref{tab:JEM_IPM_codes}} are coded with \gls{CABAC} contexts while other bits are bypassed. This coding scheme is noted by $2+3+4+5+6+6+(6 \times 16)+(7\times 19)+(8\times 26)$ giving the number of bits of the six \glspl{MPM} followed by the number of bits used to code each of the three other defined lists multiple the number of \glspl{IPM} within the list. 

In \gls{AI} coding configuration, the part of the bitstream related to \glspl{IPM} represents in average $9 \%$ and $5.2 \%$ of the whole \gls{HEVC} and \gls{JEM} bitstreams, respectively. Therefore, a slight increase in the \gls{IPM} coding efficiency would result in a significant enhancement in terms of global \gls{RD} performance.    

\begin{table}[t]
\renewcommand{\arraystretch}{1.1}
\center
	\caption{used codewords  for \glspl{IPM} in \gls{HEVC} ($2+3+3+(6 \times 32)$). The first bit in red color is coded using a \gls{CABAC} context while others are coded in "bypass mode", with no \gls{CABAC} context.}
	\label{tab:hevc_IPM_codes}
	\begin{tabular}{|l|cc|}
	\hline
		\gls{IPM}s  & Codewords & Number of bits  \\  [0.5ex]
		\hline
		\hline
		\gls{MPM}0 & \textcolor{red}{1}0 &  2 \\  [0.5ex]
		 \gls{MPM}1 & \textcolor{red}{1}10 & 3 \\  [0.5ex]
		\gls{MPM}2 &  \textcolor{red}{1}11 & 3 \\  [0.5ex]
		\hline
		   & \textcolor{red}{0}00000 & \\  [0.5ex]
		32 remaining \gls{IPM}s &  $\vdots$ & 6 \\  [0.5ex]
		  & \textcolor{red}{0}11111 & \\	
		  \hline
		  
	\end{tabular}
\end{table}

\subsection{Upper bound performance and motivation}
\label{sec:upper_bound_motivations}
In this section, we evaluate the theoretical limit (empirical entropy) of the \gls{IPM} coding using contextual information in \gls{HEVC} and \gls{JEM}. The entropy is compared to the performance of the codes defined in both \gls{HEVC} and \gls{JEM} in terms of average number of bits required to encode the \gls{IPM}. We consider a dataset composed of 59 heterogeneous video sequences for \gls{HEVC} and 300 video sequences for the \gls{JEM}. These video sequences are encoded with \gls{HM} and \gls{JEM} reference software in \gls{AI} coding configuration at 4 different \glspl{QP} $\in \{22, \, 27, \, 32, \, 37 \}$ \cite{seregin_common_2014}. The coding of \gls{IPM} is carried out with a fixed length code of 6 bits in \gls{HM} and 7 bits in the \gls{JEM}. With a fixed length code, we make sure that the signalling scheme has no impact on the \gls{IPM} selection, as all modes have the same rate cost, and thus the mode is chosen based on only its corresponding distortion. It should be noted that this configuration is used only to compute the entropy and to train our proposed model. The distribution of data used in here to train our model is not related to a specific coding solution enabling better generalisation of the proposed model.  

The entropy computation considers the modes of neighbour blocks in the conditional probability formulation $\mathbb{P}(i| U, L)$ where the modes of upper $U$ and left $L$ blocks are known at the decoder. The \gls{HEVC} and \gls{JEM} encodings construct a dataset of  $15 \times 10^6$ and $1.2\times 10^9$  samples, respectively, where each sample corresponds to the Intra block and its corresponding \gls{IPM} and contextual information, i.e. the \glspl{IPM} neighbour blocks. 

\begin{table}[t]
\renewcommand{\arraystretch}{1.1}
\center
	\caption{Used codewords  for \glspl{IPM} in \gls{JEM}: $2+3+4+5+6+6+(6 \times 16)+(7\times 19)+(8\times 26)$. the first bits in red color are coded using \gls{CABAC} contexts while others are bypassed.}
	\label{tab:JEM_IPM_codes}
	\begin{tabular}{|l|cc|}
	\hline
		\gls{IPM}s  & Codewords & Number of bits  \\  [0.5ex]
		\hline
		\hline
		\gls{MPM}0 & \textcolor{red}{10}      &  2 \\  [0.5ex]
		 \gls{MPM}1 & \textcolor{red}{110}  & 3 \\  [0.5ex]
		\gls{MPM}2 &  \textcolor{red}{1110} & 4 \\  [0.5ex]
		\gls{MPM}3 & \textcolor{red}{1111} 0 &  5 \\  [0.5ex]
		 \gls{MPM}4 & \textcolor{red}{1111} 10 & 6 \\  [0.5ex]
		\gls{MPM}5 &  \textcolor{red}{1111} 11  & 6 \\  [0.5ex]
		\hline
		   & \textcolor{red}{01} 0000 & \\  [0.5ex]
		\textit{Preferred} list  ($\times 16$)&  $\vdots$ & 6 \\  [0.5ex]
		  & \textcolor{red}{01} 1111 & \\	
		  \hline
		   & \textcolor{red}{00} 00000 & \\  [0.5ex]
		\textit{Other part 1} list ($\times 19$)  &  $\vdots$ & 7 \\  [0.5ex]
		  & \textcolor{red}{00} 10010 & \\	
		  \hline
		   & \textcolor{red}{00} 100110 & \\  [0.5ex]
		\textit{Other part 2} list  ($\times 26$) &  $\vdots$ & 8 \\  [0.5ex]
		  & \textcolor{red}{00} 111111 & \\	
		 \hline
	\end{tabular}
\end{table}

The entropy of coding \gls{IPM} using the contextual information $L$ and $U$ is computed as follows
\begin{equation}
\begin{split}
	\mathbb{H}(\gls{IPM}, L,U) = & -\sum_{L=-1}^{K-1}\sum_{U=-1}^{K-1} \mathbb{P}(L,U)  \\
	& \, \sum_{i=0}^{K-1}\mathbb{P}(i|L,U) \,  \, \log_{2} \,  \mathbb{P}(i|L,U) ,
	\label{eq:entropyHM}
\end{split}
\end{equation}
where $K$ is the number of \glspl{IPM} and $\mathbb{P}(L,U)$ is the probability of the left \gls{IPM} being equal to mode $L$ and the upper \gls{IPM} being equal to mode $U$. $\mathbb{P}(i|L,U)$ is the conditional probability of the selected \gls{IPM} being equal to $i$ in the case of the left \gls{IPM} is equal to $L$ and the above \gls{IPM} is equal to $U$ with $L, U \in \{-1, 0, 1, \dots, K-1 \}$. This conditional probability considers the contextual information from known left and upper Intra modes. The specific case of $X=-1$ refers to the context $X$ is not available. To compute the empirical entropy, we first estimate from the available prediction modes of the considered dataset the probabilities $\mathbb{P}(L,U)$ and $\mathbb{P}(i | L,U)$ with $L, U \in \{-1, 0, 1, \dots, K-1 \}^2$ and $i \in \{ 0, 1, \dots, K-1 \}$, then apply Equation
~(\ref{eq:entropyHM}) to compute the empirical entropy.

Under the \gls{HEVC} dataset, the $\mathbb{H}(\gls{IPM},L,U)$ is equal to $3.36$ bits while the \gls{HEVC} coding scheme achieves $3.86$ bits/mode on the same dataset.  Concerning the \gls{JEM}, we compute the entropy with using $L$, $U$, $BL$, $UR$ and $UL$ modes as available contextual information
\begin{equation}
\begin{split}
	\mathbb{H}&(\gls{IPM}, L,U, \dots, UL) =  \\
	 & -\sum_{L=-1}^{K-1}\sum_{U=-1}^{K-1} \dots \sum_{UL=-1}^{K-1}  \mathbb{P}(L,U, \dots, UL)  \\ 
	 & \sum_{i=0}^{K-1}\mathbb{P}(i|L,U, \dots, UL)  \log_{2} \, \mathbb{P}(i|L,U, \dots, UL).
	\label{eq:entropy}
\end{split}
\end{equation}

The entropy $\mathbb{H}(\gls{IPM},L,U, UL)$ is equal to $3.32$ bits while the \gls{JEM} code enables $3.44$ bits/mode. Table~{\ref{tab:Entropy_HEVC}} gives the entropy of the \gls{IPM} in the \gls{HEVC} and \gls{JEM} without and with using $L$, $U$ $BL$, $UR$ and $UL$ as contextual information for prediction.  By definition, the empirical entropy only provides a lower bound estimation for the amount of information contained in the signal~\cite{Paninski:2003:EEM:795523.795524}. This means that the value estimated by (\ref{eq:entropy}) is always lower than the actual information contained in the signal. Adding more data samples reduces the error made by the estimation. One possible correction is the Miller-Madow (MM) bias correction~\cite{10.2307/2237121} which corrects the entropy by adding $\frac{\tilde{m} - 1}{2B}$, where $\tilde{m}$ is an estimation of the number of bins with non-zero probability (for \gls{IPM} it is the number of possible values \{$L$, $U$, $UR$, $UL$, $BL$, \gls{IPM}\} which is equal to $67^6$) and $B$ is the number of considered samples (in our case the number of samples in the JEM dataset is $1.2\times 10^9$). The values of Miller-Madow (MM) correction are equal to 0 bits/\gls{IPM} from 0 to 2 contexts, $0.008$ bits/\gls{IPM} for 3 contexts, $0.275$ bits/\gls{IPM} for 4 contexts and $1.18$ bits/\gls{IPM} for 5 contexts. Therefore, the empirical entropy computed with more than four contexts is not reliable since the error is high compared to the computed entropy. 

The difference between the computed entropy and the performance of \gls{HEVC} \glspl{IPM} coding, estimated to $0.5$ bits/mode in Table~{\ref{tab:Entropy_HEVC}}, clearly shows that there is a room for improvement of the \gls{IPM} coding scheme. This motivates the design of more efficient codes that outperform the existing codes under similar coding/decoding complexities. However, the \gls{JEM} performance is close to the estimated entropy and thus the objective of the proposed approach is to derive a simple coding scheme with similar coding efficiency and complexity than the \gls{JEM}. Finally, the developed approach would be generic in the way that it can be applied to any other discrete predictive coding using contextual information such as other video syntax elements including motion vectors.

\begin{table}[t]
\center
\renewcommand{\arraystretch}{1.1}
	\caption{Entropy and Miller-Madow (MM) correction in bits using $L$, $U$, $UR$, $BL$, $UR$ and $UL$ contextual information in HM and \gls{JEM} datasets.}
	\label{tab:Entropy_HEVC}
\begin{tabular}{|l|cc|cc|}
	\hline
		  Codec &    \multicolumn{2}{c|}{HEVC} & \multicolumn{2}{c|}{JEM}   \\  [0.5ex]
		   & Entropy & MM &  Entropy & MM \\  [0.5ex]
		\hline
		\hline
		 $\mathbb{H}(\gls{IPM})$ & $4.73$ &$0$ &  $4.66$ & $0$ \\  [0.5ex]
		 $\mathbb{H}(\gls{IPM} , L)$ &  $3.91$ & $0$ & $3.79$ &  $0$ \\  [0.5ex]
		 $\mathbb{H}(\gls{IPM} , L, U)$ & $3.36$& $0$ &  $3.37$&  $0$ \\  [0.5ex]
		 $\mathbb{H}(\gls{IPM} , L, U, UL)$ & -- & -- & $3.32$  & $0.008$\\  [0.5ex]
		 $\mathbb{H}(\gls{IPM} , L, U, UR, UL)$ & -- & -- &  $3.19$ & $0.275$ \\  [0.5ex]
		$\mathbb{H}(\gls{IPM} , L, U, BL, UR, UL)$ & -- & -- & $2.99$ & $1.18$ \\  [0.5ex]
		  \hline
		 Anchors & \multicolumn{2}{c|}{$\textcolor{red}{3.86}$} & \multicolumn{2}{c|}{$\textcolor{red}{3.44}$}  \\
		  \hline
	\end{tabular}
	\end{table}

\section{Predictive coding model of IPM} 
\label{model}
In this section the predictive coding using contextual information is modelled in three main steps: {\it prediction}, {\it clustering} and {\it coding}. For each step, examples with the \gls{IPM} coding schemes used in \gls{HEVC} and \gls{JEM} are provided. 
\subsection{Prediction}
\label{sec:lists}
Prediction enables to reduce the entropy of the information to encode with using already known side information. In video coding continuous prediction is used to predict the pixels based on previously decoded pixels within the same frame for Intra prediction or previously decoded frames for Inter prediction. The prediction of \gls{IPM} belongs to the discrete (non-continuous) prediction. The prediction of \gls{IPM} is carried out with contextual information including the \gls{IPM} of already coded neighbour blocks as well as some specific numerical modes with high probability to be selected. These \glspl{IPM} used as predictors are referred in this paper as {\it labels}. 

\gls{HEVC} defines 7 \textit{labels} for the prediction of the \gls{IPM} which are based on two contextual information: \gls{IPM} of the left block ($L$) and \gls{IPM} of the upper block ($U$). Four \textit{labels} in \gls{HEVC} are based on these two contextual information: $L$, $U$, two angular neighbours of mode $L$ ($L+1$ and $L-1$) as well as three numerical \glspl{IPM} : 0 (planar), 1 (DC) and 26 (vertical direction). The \gls{JEM} defines a set 15 \textit{labels} from 5 contextual information including \glspl{IPM} of the left block ($L$, $L\pm1$), the upper block ($U$, $U\pm1$), below left block  ($BL$, $BL\pm1$), upper right block ($UR$, $UR \pm 1$), upper left block  ($UL$, $UL\pm1$) and six numerical \glspl{IPM}: Planar, DC, $50$ (vertical), $18$ (horizontal), $34$ (diagonal) and $2$ (the other diagonal). In total, the \gls{JEM} defines a larger set of 21 { \it labels} instead of only seven in \gls{HEVC}.

\subsection{Clustering}
Clustering is the step that enables to distinguish data following its probability distribution. Better performance will usually be achieved if a prediction is flexible or robust in the sense of being able to fit the local statistical behaviour of the signal~\cite{Gersho_vector_1992}. In other words, a smart prediction should adapt to the source at hand. This enables better conditioning and reduces the global entropy of the system~\cite{855427}. To perform clustering, a set of \textit{tests} is defined to build a binary-tree where the leaves vertices represent the different clusters of {\it labels} and the nodes vertices correspond to {\it tests}.

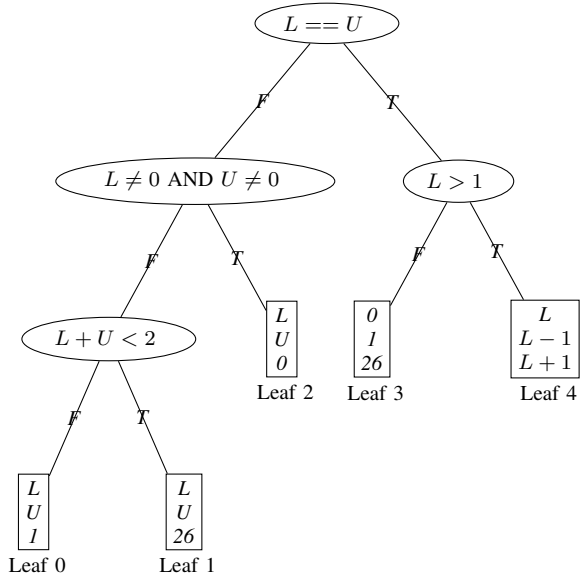
\begin{figure}
\centering
\begin{footnotesize}
\tikzstyle{level 1}=[level distance=2.1cm, sibling distance=3.5cm]
\tikzstyle{level 2}=[level distance=2.1cm, sibling distance=2.3cm]
\tikzstyle{level 3}=[level distance=2.3cm, sibling distance=2cm]
\tikzstyle{bag} = [text centered, draw, ellipse, align=center]
\tikzstyle{end} = [minimum width=3pt, draw, align=center]
\begin{tikzpicture}[grow=down]
\node[bag] { {$L==U$} }
child
{
	node[bag] { {$ L\neq 0 $} AND  {$ U\neq 0 $} }
	child
	{
		node[bag] { {$ L+U<2 $} }
		child
		{
			node[end, label=below: { Leaf 0 }]
				{ \textit{L} \\ \textit{U} \\ \textit{1} }
			edge from parent
				node[] { \textit{F} }
		}
		child
		{
			node[end, label=below: { Leaf 1 } ]
				{ \textit{L} \\ \textit{U} \\ \textit{26}}
			edge from parent
				node[] { \textit{T} }
		}
		edge from parent
			node[] { \textit{F} }
	}
	child
	{
		node[end, label=below: { Leaf 2 }]
			{ \textit{L} \\ \textit{U} \\ \textit{0}}
		edge from parent
			node[] { \textit{T} }
	}
	edge from parent
		node[] { \textit{F} }
}
child
{
	node[bag] { {$ L>1 $} }
	child
	{
		node[end, label=below: { Leaf 3 }]
			{  \textit{0} \\ \textit{1} \\ \textit{26}}
		edge from parent
			node[] { \textit{F} }
	}
	child
	{
		node[end, label=below: { Leaf 4 }]
			{  \textit{L} \\ {$ L-1$} \\ {$L+1$} }
		edge from parent
			node[] { \textit{T} }
	}
	edge from parent
		node[] { \textit{T} }
}
;
\end{tikzpicture}
\end{footnotesize}
\caption[\gls{HEVC} Decision tree]{\gls{HEVC} binary-tree used to create the \gls{MPM} list. The code used is $2+3+3+(6 \times 32)$.}
\label{fig:hevc_tree}
\end{figure}

In \gls{HEVC}, four different {\it tests} are defined to separate the {\it labels} into five clusters. Fig.~{\ref{fig:hevc_tree}} illustrates the binary-tree used in \gls{HEVC} to build these five clusters of { \it labels} represented by the leaves vertices and the used {\it tests} given by the labelled vertices. The {\it tests} leverage the two contextual \textit{labels}: $L$ and $U$ while the clusters of the three { \it labels} represent all possible \glspl{MPM} lists. The used \textit{tests} focus mainly on finding whether the two modes $L$ and $U$ are equal and whether they are angular (i.e. $L > 1$) or not.

The \gls{JEM} defines more {\it tests} to build the binary-tree with leaves vertices representing the different clusters of six {\it labels}. We do not give here the binary-tree of the \gls{JEM} since it consists of 460 leaves and 196 different labellings composed of a list of 21 available {\it labels}. Moreover,  the predictive coding of \gls{IPM} in the \gls{JEM}  was not designed as a decision tree. In fact, the \gls{JEM} uses a dynamic list to construct the \glspl{MPM} list based on an ordered list of \textit{labels}: the \textit{labels} are added to the \glspl{MPM} list while checking, for each additional \textit{label}, whether the corresponding \gls{IPM} is already in the \glspl{MPM} list or not.

\begin{figure}[ht]
	\centering
	\subfloat[IPM probabilities $L=10$]{\includegraphics[width= 0.5\linewidth]{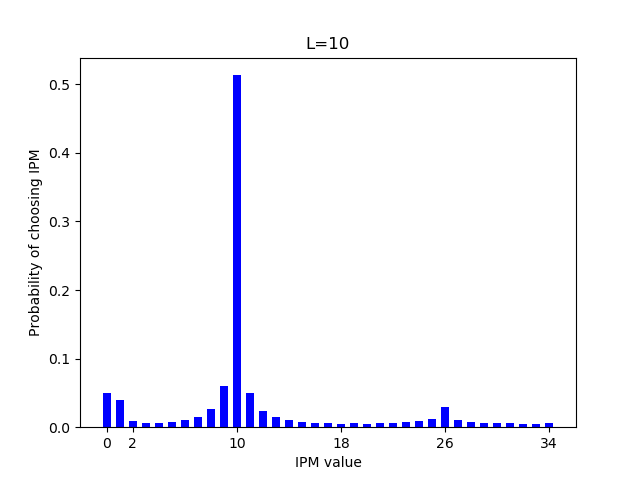}\label{fig:intra_mode_proba10}}
	\subfloat[IPM probabilities $L=18$]{\includegraphics[width= 0.5\linewidth]{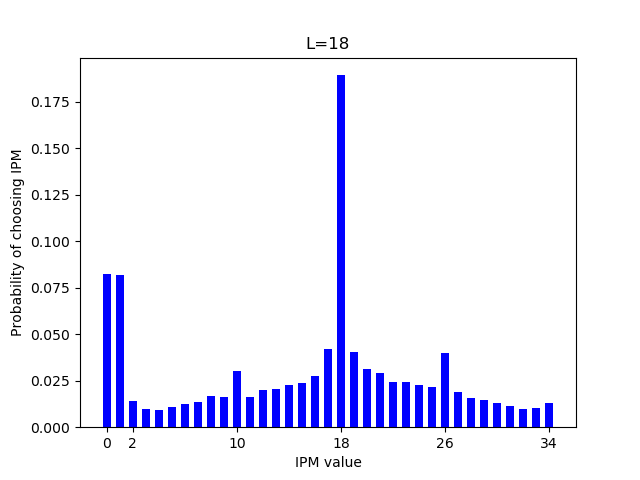}\label{fig:intra_mode_proba18}} 
	
	\subfloat[IPM probabilities $L=26$]{\includegraphics[width= 0.5\linewidth]{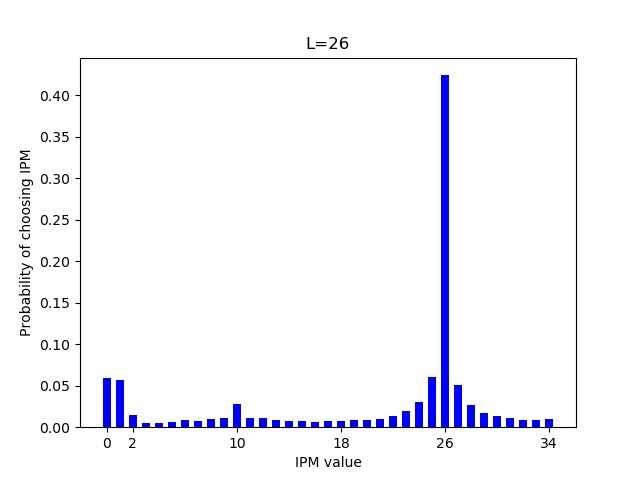}\label{fig:intra_mode_proba26}} 
	\subfloat[IPM probabilities $L=30$]{\includegraphics[width= 0.5\linewidth]{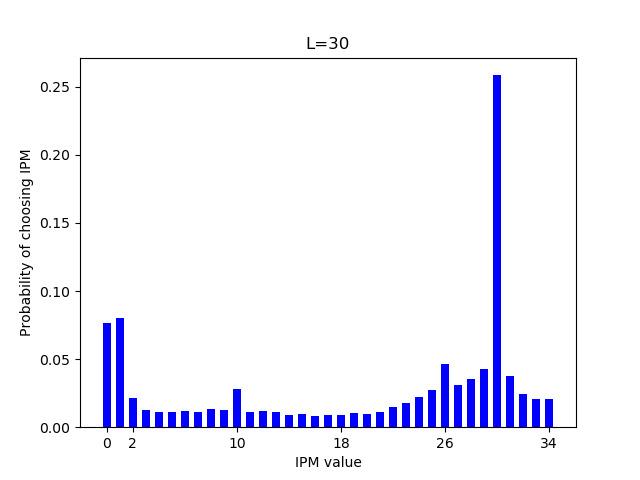}\label{fig:intra_mode_proba30}} 
	\caption{Probability distribution  $\mathbb{P}(i|L)$ of choosing an IPM $i \in \{0, 1, \dots, 35 \} $ when the IMP of the left neighbour block $L$ is:  (a) $L=10$, (b) $L=18$, (c) $L=26$, (d) $L=30$. The data is from the HM training set encoded in All Intra configuration at 4 QPs}.
	\label{fig:intra_mode_proba}
\end{figure}

\subsection{Coding}
The coding step defines the codes used to signal the selected \gls{IPM} at the encoder to the decoder. An efficient coding would allocate a short length code to the most probable modes and longer length code for the rest of modes. After the construction of the binary tree by the clustering step, the coding step allocates a specific code for the \glspl{IPM} included in the different lists. One code for all the \gls{MPM} lists (leaves vertices or clusters) can be used, or multiple-code can be used, one specific code for each \glspl{MPM} list (one specific code by leave vertex or cluster).

In \gls{HEVC}, \gls{VL} code is used to encode all \glspl{IPM} within the \glspl{MPM} list (one \gls{VL} code for all leaves vertices) and one \gls{FL} code is used to encode the remaining \gls{IPM} not present in the selected \glspl{MPM} list. The codes used in \gls{HEVC} are depicted in Table~{\ref{tab:hevc_IPM_codes}}. As for \gls{HEVC}, \gls{JEM} uses one \gls{VL} code to all possible modes in \glspl{MPM} list. Moreover, the \gls{JEM} defines three other \gls{FL} codes to encode the \glspl{IPM} in the three other lists: \textit{Preferred}, \textit{Other part 1} and \textit{Other part 2} as depicted in Table~\ref{tab:JEM_IPM_codes}. 

In this section we have modelled the predictive coding of \gls{IPM} in three steps: { \it prediction}, { \it clustering} and { \it coding}. We have also defined the \textit{labels}, \textit{tests} and \textit{codes} used to encode the \glspl{IPM} in both \gls{HEVC} and \gls{JEM}. In the next sections, we improve these three steps to enhance the coding performance of both \gls{HEVC} and \gls{JEM} with defining additional \textit{labels}, \textit{tests}, \textit{codes} and thus new binary trees to reach higher coding performance.

\section{Efficient coding schemes of IPMs in HEVC and JEM} 
\label{HMJEM}

\subsection{Improved predictive coding}
\label{sec:improved_predictive_coding}
In this section the three predictive coding steps for \gls{IPM} coding in both \gls{HEVC} and \gls{JEM} are improved. The improvements proposed for \gls{HEVC} and \gls{JEM} are investigated separately since the same level of complexity with the anchor codes defined in \gls{HEVC} and \gls{JEM} is targeted, respectively. 

\subsubsection{HEVC} 
To improve the prediction coding of \gls{IPM} in \gls{HEVC}, we first introduce additional \textit{labels} to extend the list of \textit{labels} to better leverage contextual information. The new list of \textit{labels} includes the 7 \textit{labels} defined in \gls{HEVC} and 23 new \textit{labels} resulting in a new list of 30 contextual \textit{labels} and 5 numerical modes. Fig.~\ref{fig:intra_mode_proba} gives the conditional probability distribution of the IPM $\mathbb{P}(i|L)$ when the mode of the left neighbour block is equal to 10, 18 26 and 30.  We can notice from this figure that in addition to the DC and PL modes, the probabilities of modes around the $L$ mode are also high compared to the rest of modes. This motivates including up to six modes around both left ($L$) and ($U$) modes as {\it labels}.   
The new {\it labels} are selected from the {\it labels} leading to highest conditional probabilities $\mathbb{P}(i|L)$, $\mathbb{P}(i|U)$ with $i \in \{ L, U, L\pm1, U \pm 1, L\pm2, U \pm 2, L\pm3, U \pm 3, \dots \}$ as well as the most used numerical modes. The selected labels are $labels \in \{L, \, U, \, L\pm1, \, 0, \, 1, \, 26, \, U\pm1,  \, L\pm2, \, L \pm 3, \, U \pm 2, \, U \pm 3, \,  \min(L, \, U), \max(L, \, U), \,  \min(L, \, U)\pm1, \, \max(L, \, U)\pm1, \, \min(L, \, U)\pm2, \, \max(L, \, U)\pm2, \, \min(L, \, U) \pm 3, \, \max(L, \, U) \pm 3,   |1-min(L,U)|, \, (L+U)/2, \, 18, \, 2  \}$. 

To leverage these new \textit{labels}, new \textit{tests} are also added to the four \gls{HEVC} \textit{tests} set:   $tests \in \{ L==U, \,
|L-U|<2, \,  |L-U|==2 , \, \min(L,U)>1, \, \min(L,U)<1, \, \max(L,U)<2 , \, L+U<2 , \,|L-10|<3 , \,|L-26|<3 , \, |L-18|<3 , \, |U-10|<3 , \, |U-26|<3 , \, |U-18|<3 , \, L<2 \}$. These new \textit{tests} are then derived to improve the clustering by finding more similarities. For example, the test $|L - U| < 2$ is added as an extension to the test $L == U$. This new test, when $L$ and $U$ are angular, allows detecting whether the two modes are really close to each other. Other new \textit{tests} check the proximity of the vertical, horizontal and diagonal directions (resp. 26, 10 and 18). \textit{Tests} using the \textit{labels} $\min(L, U)$ and $\max(L, U)$ are also proposed. The \textit{tests} $\min(L, U) > 1$ and $\max(L, U) < 2$ are added to determine whether the modes are both angular or non-angular ($PL$ or $DC$). 

The last step consists in defining the used  codes at each cluster. These codes will also determine the number of \glspl{MPM} within the cluster. A list of \gls{VL} \textit{codes} is created, including all codes for clusters of 3 to 7 \glspl{MPM}. These codes represent all possible \gls{VL} codes with 3 to 7 bits and \gls{FL} codes for the rest of symbols. This results in 4 different codes using 3 \glspl{MPM}, 8 using 5 \glspl{MPM} and 43 using 7 \glspl{MPM} for a total of 55 different codes. We consider in this work two configurations: the first one, similar to \gls{HEVC} and \gls{JEM}, that uses only one code for all clusters while the second configuration that defines different codes for the clusters.   

This new set of codes enables a new theoretical limit on the best achievable cost. The value of $3.45$ bits/IPM is the entropy of the data. To achieve such performance the system would need to use an entropy encoder for all bits of the coding scheme. Moreover, it would require the use of all possible coding schemes for 35 symbols (for which there are 108,861,148 possibilities~\cite{Handbook_of_Integer}), and not only the 55 codes described above. This list of codes motivates the computation of another theoretical limit which considers only the 55 selected codes. 

Equation~(\ref{eq:entropyHM}) is modified to model this description and compute what we define as the empirical code-based entropy. The term $-\log_2  \mathbb{P} \left (i|L, U \right )$ in (\ref{eq:entropyHM}) is replaced with the used code-length to encode the \gls{IPM} $i$ knowing the \glspl{IPM} of left and upper neighbour blocks $L$ and $U$, respectively. 
\begin{equation}
\begin{split}
	&\mathbb{H}_{CB}(\gls{IPM},L,U) =  \\
	 &\sum_{L=-1}^{K-1}\sum_{U=-1}^{K-1}  \,  \sum_{i=0}^{K-1}\mathbb{P}(i|L,U) \,  \, CodeL( i|L,U ).
	\label{eq:code_based_entropyHM}
\end{split}
\end{equation}

This results in a system with $K \times K$ clusters (with $K=35$ in HEVC and $67$ for JEM), each cluster uses the most efficient code among the 55 codes. In \gls{HEVC}, the code-based entropy achieves a performance of $3.52$ bits/\gls{IPM}. The signalling scheme giving this value can be created. However, it cannot be implemented because of the high number of clusters. In practice, implementing such a system would require having a table of 1225 entries, each one contains the \textit{code} and \textit{labels} to use, which is not acceptable for real time applications. This value asserts that the created systems can reduce the average cost of \gls{IPM} signalling in \gls{HEVC} when the chosen \textit{tests} and \textit{labels} allow to reach this limit for a small number of clusters. 

It should be noted that all these \textit{labels}, \textit{tests} and \textit{codes} will be considered to construct the binary tree of our signalling scheme while only few selected \textit{labels}, \textit{tests} and \textit{codes} will be used within the final binary tree. The selection is performed by a genetic algorithm described in the next section.   


\begin{table}[h]
\renewcommand{\arraystretch}{1.1}
\center
\caption{Code-based entropy (CBE) and Miller-Madow (MM) correction  in bits using $L$, $U$, $UR$, $BL$, $UR$ and $UL$ contextual information on HM and \gls{JEM} datasets.}
	\label{tab:Entropy_HEVC_JEM}
\begin{tabular}{|l|cc|cc|}
	\hline
Codec &     \multicolumn{2}{c|}{HEVC} & \multicolumn{2}{c|}{JEM}   \\  
&   CBE & MM &  CBE  & MM   \\  
\hline \hline
$\mathbb{H}_{CB}(\gls{IPM})$ &   $4.84$ & $0$  &  $4.70$ &   $0$   \\ 
$\mathbb{H}_{CB}(\gls{IPM} , L)$ & -   & -& $3.85$&  $0$  \\  
$\mathbb{H}_{CB}(\gls{IPM} , L, U)$ & $3.52$   & $0$ & $3.43$ & $0$   \\ 
$\mathbb{H}_{CB}(\gls{IPM} , L, U, UL)$ & -   & - & $3.39$ & $0.008$   \\
$\mathbb{H}_{CB}(\gls{IPM} , L, U, UR, UL)$ &   - &- &$3.29$ &   $0.275$ \\  
$\mathbb{H}_{CB}(\gls{IPM} , L, U, BL, UR, UL)$&  -  & - & $3.10$&  $1.18$  \\  
\hline
Anchors &    \multicolumn{2}{c|}{$\textcolor{red}{3.86}$} &
\multicolumn{2}{c|}{$\textcolor{red}{3.44}$}     \\  
\hline
\end{tabular}
\end{table}

\subsubsection{JEM} 
Three new lists of \textit{labels}, \textit{tests} and \textit{codes} are defined for the \gls{JEM}. The list of \textit{labels} contains 57 \textit{labels} based on three modes from contextual information  $L$, $U$ and $UL$, derived \textit{modes} from these three modes and numerical modes. We also define 41 \textit{tests} to leverage the considered 57 \textit{labels} and build the  binary tree. We define \textit{codes} to encode the clusters of 3 to 9 \glspl{IPM}. We use 4, 8, 47 and 89 \textit{codes} for lists of 3, 5, 7 and 9 \glspl{MPM}, respectively, resulting in 148 \textit{codes}. The new entropy with this set of codes can be computed from (\ref{eq:entropy}) by replacing the term  $-\log_{2} {\left ( \mathbb{P} \left (i|L,U, \dots, UL \right ) \right )}$ by the code-length of the \gls{IPM} $i$ ($i \in \{0, \, 1,\, ..., \, 66 \}$) knowing the modes of left, upper and upper left neighbour blocks are equal to $L$, $U$ and $UL$, respectively. 

The code-based entropy $\mathbb{H}_{CB}$ computed for the \gls{JEM} dataset with $L, U$ and $UL$ as contextual information is equals to $3.40$ bits/\gls{IPM}. We can notice that this code-based entropy is close to the \gls{JEM} performance of $3.44$ bits/\gls{IPM}. Therefore, there is less gain to reach with the \gls{JEM} ($<0.1$ bits/\gls{IPM}) than for \gls{HEVC} $0.34$ bits/\gls{IPM}. The objective for the \gls{JEM} is then to achieve the entropy performance of $3.4$ bits/\gls{IPM} by a simple coding scheme obtained with the proposed method. Finally, Table~{\ref{tab:Entropy_HEVC_JEM}} summarises the code-based entropy of the \gls{HEVC} and \gls{JEM} with using the considered codes and contextual information $L$, $U$, $UR$, $BL$, $UR$ and $UL$.

\subsection{Optimal search and limitations}
The objective of this work is to derive the most efficient signalling scheme of \glspl{IPM}. Therefore, we need to assert the efficiency of the proposed approach. This latter can be extended to search for the best signaling scheme of other video syntax elements. 

We separate the problem into three parts: 1) how to choose the best performing  \textit{labels} and \textit{tests}, 2) how to perform the most efficient prediction on a leaf from the available set of \textit{labels} and 3) how to choose the best clustering from the available set of \textit{tests}.

\subsubsection{Choosing labels and tests}

The \textit{labels} and \textit{tests} are used for prediction and clustering, respectively. In fact, some \textit{labels} cannot be used without the proper \textit{tests}. For instance, \textit{labels} $L$ and $U$ cannot be used together in the same cluster if no test such as $L = U$ distinguishes equality between these two modes from other cases.

When choosing \textit{tests} and \textit{labels}, it is hard to assert that the solution is optimal. There are as many possible \textit{tests} as there are functions that use the available contexts to separate the space into two clusters. We have also several \textit{labels} defined by functions combining the available contexts. The possible functions cannot all be evaluated to assert that the best performing \textit{tests} and \textit{labels} have been selected. Moreover, we restrict the study to simple \textit{tests} and simple \textit{labels} as we do not want to use \textit{tests} that combine several \textit{tests} in the form "$L \neq U~and~L < 18~and~U > 26$". However, we can try to simulate what the best performing \textit{labels} with the best clustering would look like. We can then select our \textit{tests} and \textit{labels} based on this system, to ensure that the best performing \textit{tests} and \textit{labels} are included in the considered sets. 

The upper bound theoretical limit is computed by the code-based entropy described in Section~\ref{sec:improved_predictive_coding}. It provides the best signalling scheme that can be implemented as it computes the cost of the signalling scheme that uses all possible individual clusters (i.e. the maximum number of clusters). We propose the evaluation of a new limit, similar to the code-based entropy, but for a given (and small) number of clusters. This way a limited number of codes is used and the cases with similar statistics are merged into the same cluster.

No assumptions are made on the available \textit{tests} and \textit{labels}. While the resulting system can be implemented, and consists of a real signalling scheme, the limitation on its implementation is the same as with the code-based entropy: implementation would require a complete table of all possible values of the contexts containing the corresponding cluster and the used codes. This shows the performance that can be achieved with perfect \textit{tests} and \textit{labels} but with a limited number of clusters. Therefore, the perfect \textit{tests} and \textit{labels} configuration would rely on a look-up table to derive the corresponding cluster whatever the value of the contextual information (ie. L and U modes in the case of \gls{HEVC}). This configuration relying on look-up table is not suitable in our system for complexity and memory overheads, while it is used to assess the efficiency of the selected {\it tests} and {\it labels}.      

A genetic algorithm~\cite{MCCALL2005205} is used to create this signalling scheme. Genetic algorithms are adaptive heuristic search algorithm based on the evolutionary ideas of natural selection and genetics~\cite{pal2016genetic}. They also enable intelligent and efficient exploitation of random search.

In this experiment, the $K \times K$ possible context values ($U$ and $L$) are initialized with a random cluster value.
From this parent clustering, some children are generated with small changes on the parent: some context values are affected by other clusters. The best result in terms of rate cost computed by (\ref{eq:code_based_entropyHM}) of all current parents and children is always kept as a parent for the next iteration, as well as some other randomly selected children. There is no proof that the clustering generated from this genetic search is optimal. The goal is to be able to observe the data, to see which statistics are similar and thus should use a similar \textit{code}. When using only two contexts, we can observe the results on a plan which is useful to select the appropriate \textit{tests}. 

\subsubsection{Labelling selection}

This task searches for the best compatible $M$ {\it labels} used at each leaf, with  $M$ being the number of \glspl{MPM}. The \textit{labels} are compatible on a given leaf if whatever the values of the contexts in this leaf, a couple of \textit{labels} will always correspond to two different \glspl{IPM}. Otherwise, the signalling scheme will not be valid resulting in two codewords encoding the same \gls{IPM}. As there are as many codewords as \glspl{IPM}, one \gls{IPM} cannot be encoded and correctly transmitted to the decoder. 

\begin{figure}[t]
	\begin{scriptsize}
	\begin{tabular}{|l|ccccccc|}
\hline
	Label & $L$ & $U$ & 0 & 1 & 26 & $L+1$ & $L-1$	\\
	\hline \hline
	Probability & \multirow{2}{*}{35.9\%} & \multirow{2}{*}{34.8\%} & \multirow{2}{*}{11.2\%} & \multirow{2}{*}{10.6\%} & \multirow{2}{*}{6.5\%} & \multirow{2}{*}{4.3\%} & \multirow{2}{*}{4.1\%} \\
	($\times 100 \%$)  & & & & & & & \\
	\hline
	\end{tabular}
	\end{scriptsize}
	\centering
	\includegraphics[width=\textwidth/2]{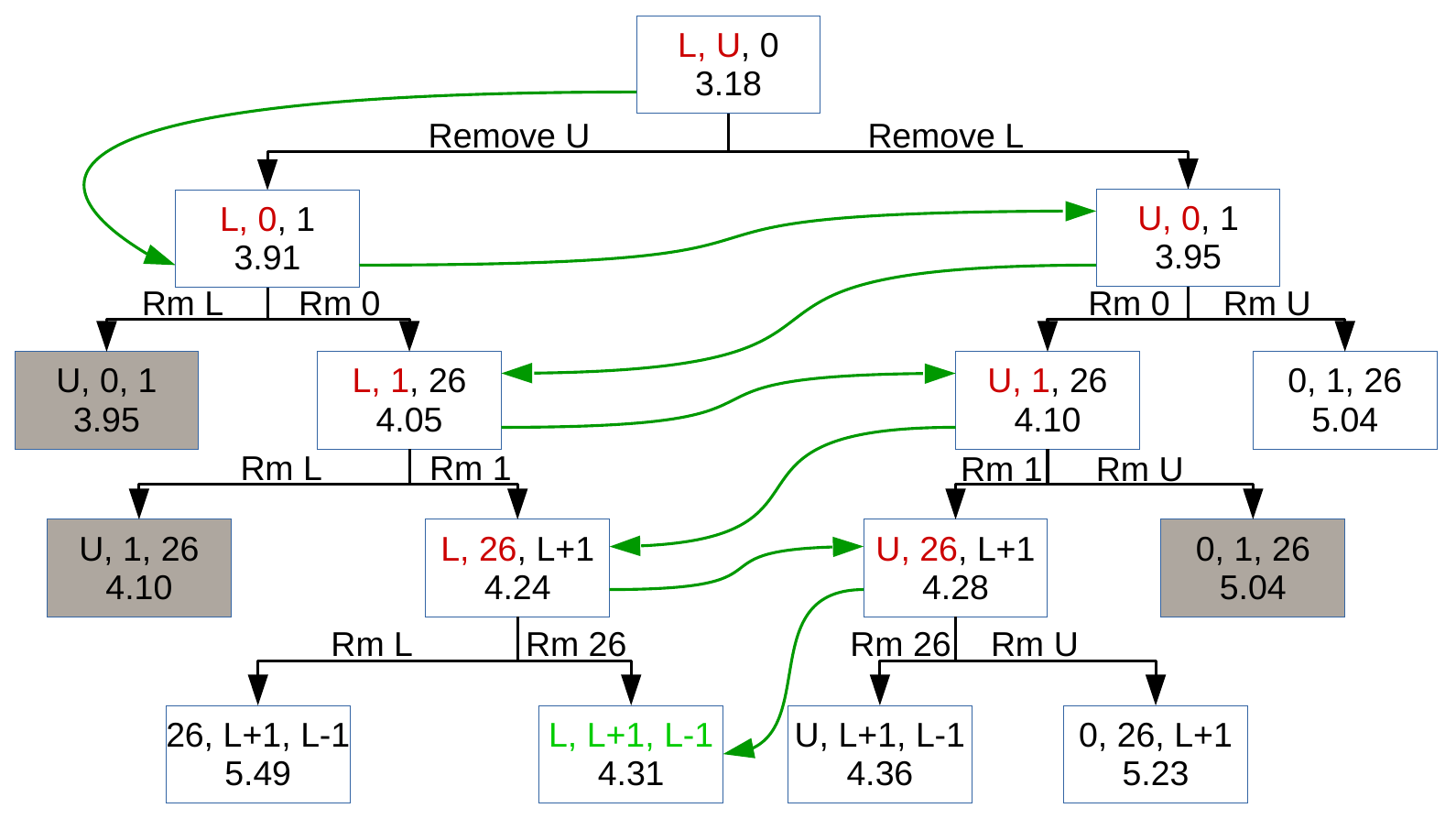}
	\caption{An example of the path followed by the selection algorithm to find the best labelling. In this case the available labels are the ones used by \gls{HEVC}. The code used to evaluate the labels is $2+3+3+(6\times 32)$. Conflicts are highlighted in red, and the cost in bits/\gls{IPM} is shown for each tested labelling. The path followed by the algorithm is shown in green. The gray blocks are not evaluated as they are  already explored.}
	\label{fig:label_search_example}
\end{figure}

An exhaustive search would result in testing all possible combinations of \textit{labels} for a given number of \glspl{MPM} to see which ones provide valid combinations. Performing this search on all leaves would be extremely costly in terms of computational complexity: to choose 7 \glspl{MPM} out of 35 labels there are ${35}\choose{7}$ $= 6,724,520$ combinations, while this search needs to be done for every available code. We propose to use an alternative approach enabling to provide the most efficient labelling while decreasing the exhaustive search complexity.  
The proposed algorithm consists in performing a greedy search carried out on all the \textit{labels}. In a greedy search, the goal is to maximize the gain at each step.
For the {\it label} search it means always starting from the available labelling with the lowest cost. 
Fig.~{\ref{fig:label_search_example}} gives an example the label search algorithm.

The algorithm starts with the best possible list of \textit{labels} by selecting the most probable \textit{labels} and puts them in the \glspl{MPM} list. The list is then checked for validity on the leaf by asserting that whatever the context values, all \textit{labels} are compatible (i.e. all \textit{labels} are always two by two different). In the case of incompatibility, two new lists are created: each with one of the two conflicting \textit{labels} removed and replaced by the following {\it label} with the highest probability in the list. The two newly created lists are added to a queue, and the next list to test is the one with the smallest cost from the queue. The cost in  bits/\gls{IPM} is computed by~(\ref{eq:code_based_entropyHM}).

It can be shown that all labellings are tested in increasing cost order, and that the best labelling is always added to the queue, and thus tested by the algorithm. Therefore, the first labelling for which all \textit{labels} are compatible is selected as the most efficient one.
\subsubsection{Clustering selection}

Two different algorithms are studied for the clustering: one for \gls{HEVC} and one for the \gls{JEM}. We target for \gls{HEVC} a maximum of 8 different clusters and binary-tree with depth 4. Therefore, we can afford performing an exhaustive search over all possible combinations of \textit{tests}.

However, the \gls{JEM} coding scheme requires higher number of clusters and  deeper binary-tree, so this exhaustive search becomes complex, especially when searching binary-trees with a depth higher than 4. Therefore, the \textit{tests} will be combined using a genetic search algorithm. This genetic algorithm starts from a binary-tree with random selection of {\it tests} and, at each iteration, randomly changes some tests. The best performing configuration is kept for the next iteration as well as some of the created children.

\subsection{Optimal clustering search validation}

\begin{figure}[t]
	\centering
	\subfloat[Perfect labels and tests (3.53 bits/IPM)]{\includegraphics[width= 0.5\linewidth]{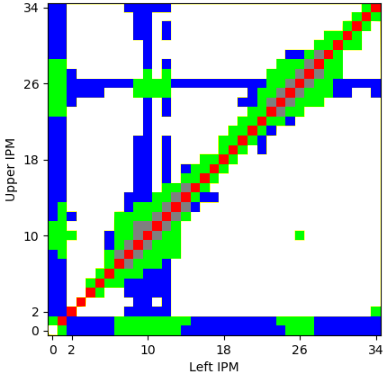}\label{fig:clustering-perf}} 
	\subfloat[Labels from the new list of labels and perfect tests (3.60 bits/IPM)]{\includegraphics[width= 0.5\linewidth]{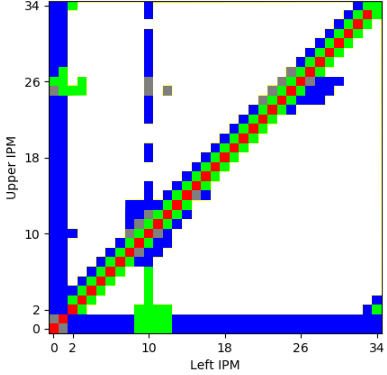}\label{fig:clustering-tests}} 
	\caption{Clustering results according to the left and upper IPM. The five clusters are illustrated in different colors.}
	\label{fig:clustering}
\end{figure}


Fig.~{\ref{fig:clustering}} shows the results of the genetic clustering for 5 clusters (similar to \gls{HEVC}) using (a) perfect \textit{labels} and \textit{tests}, (b) \textit{labels} from the proposed list of \textit{labels} and perfect \textit{tests}. These two configurations enable coding efficiencies on the training set of 3.53 and 3.60  bits/IPM, respectively. The 5 clusters are illustrated with different colour codes with respect to the left $L$ and upper $U$ \glspl{IPM}. This clustering asserts whether the selected \textit{labels} and \textit{tests} are sufficient or we need to add new ones. The red cluster is used when $L$ and $U$ are equal and the white cluster represents the case where $L$ and $U$ are not equal and different from 0 and 1. The rest of the clustering is harder to interpret but there seem to be differences when L or U are not both angular; when L and U are close but not equal; when L is close to 10 (horizontal mode); when U is close to 26 (vertical mode). Those last three cases are not represented by any leaf in the \gls{HEVC} binary-tree. As all those \textit{tests} are available in our list of \textit{tests}, we can assume that our \textit{tests} are diverse enough.

We can notice that clustering in Fig.~\ref{fig:clustering-tests} is close to the one using perfect labels illustrated in Fig.~\ref{fig:clustering-perf}. This asserts that the proposed {\it labels} is close to the perfect labels. However, this clustering cannot be reproduced by a simple tree since the perfect test is supposed. This means that the performance of this signalling can only be obtained with a high number of leaves. However, if the tests and labels are efficient enough to reproduce this clustering, this would mean having a tree that uses 3.60 bits/IPM, which would be a significant improvement over the HEVC signalling scheme, which uses 3.86 bits/IPM.

\begin{table}[H]
\renewcommand{\arraystretch}{1.1}
\centering
\caption{Test video sequences}
\label{table::seq_ctc}
\begin{tabular}{| l | l c c c |}
\hline
\multirow{2}{*}{Class} & \multirow{2}{*}{Name} & \multirow{2}{*}{Resolution} & Frame  & Number \\  
 &  &  & rate (fps) & of frames \\  
\hline \hline
\multirow{4}{*}{A1} & \textit{CampfireParty} & \multirow{4}{*}{3840$\times$2160} & 30 & 300 \\  
 & \textit{Drums} &  & 100 & 300 \\  
\scriptsize{\gls{JEM}} & \textit{Tango} &  &  60 & 294 \\  
 & \textit{ToddlerFountain} &  & 60 & 300 \\  
 \hline
 \multirow{4}{*}{A2} & \textit{CatRobot} & \multirow{4}{*}{3840$\times$2160} & 60 & 300 \\  
 & \textit{DaylightRoad} &  & 30 & 300 \\  
\scriptsize{\gls{JEM}} & \textit{RollerCoaster} &  & 60 & 300 \\  
 & \textit{TrafficFlow} &  & 60 & 300 \\  
 \hline
\multirow{4}{*}{A} & \textit{NebutaFestival} & \multirow{4}{*}{2560$\times$1600} & 30 & 150 \\  
 & \textit{PeopleOnStreet} &  & 30 & 150 \\  
\scriptsize{\gls{HEVC}} & \textit{SteamLocomotiveTrain} &  & 30 & 150 \\  
 & \textit{Traffic} &  & 30 & 150 \\  
 \hline
\multirow{5}{*}{B} & \textit{BasketballDrive1} & \multirow{5}{*}{1920$\times$1080} & 24 & 240 \\  
 & \textit{BQTerrace} &  & 24 & 240 \\  
 & \textit{Cactus} &  & 50 & 500 \\  
 & \textit{Kimono1} &  & 50 & 500 \\  
 & \textit{ParkScene} &  & 60 & 600 \\  
 \hline
\multirow{4}{*}{C} & \textit{BasketballDrill} & \multirow{4}{*}{832 $\times$ 480}  & 24 & 240 \\  
& \textit{BQMall} &  & 24 & 240 \\  
& \textit{PartyScene} &   & 24 & 240 \\  
& \textit{RaceHorses} &   & 24 & 240 \\  
\hline
\multirow{4}{*}{D} & \textit{BasketballPass} & \multirow{4}{*}{416 $\times$ 240}  & 24 & 240 \\  
 & \textit{BlowingBubbles} &   & 24 & 240 \\ 
 & \textit{BQSquare} &  & 24 & 240 \\  
 & \textit{RaceHorses} &   & 24 & 240 \\  
 \hline
\multirow{3}{*}{E} & \textit{FourPeople} & \multirow{3}{*}{1280 $\times$ 720}  & 24 & 240 \\  
& \textit{Johnny} &   & 24 & 240 \\  
& \textit{KristenAndSara} &   & 24 & 240 \\  
\hline
\multirow{4}{*}{F} & \textit{BasketballDrillText} & 832 $\times$ 480  & 24 & 240 \\  
 & \textit{ChinaSpeed} & 1024 $\times$ 468 & 24 & 240 \\  [0.5ex]
\scriptsize{\gls{HEVC}} & \textit{SlideEditing} & 1280 $\times$ 720 & 24 & 240 \\  
 & \textit{SlideShow} & 1280 $\times$ 720 & 24 & 240 \\  
\hline
\end{tabular}
\end{table}

The results of this genetic clustering are similar on the \gls{JEM} but harder to illustrate as more than two contexts are used. The efficiency of the proposed solution for the \gls{JEM} signalling scheme also needs to be assessed by comparing the trees of small depths created with the genetic algorithm to the best trees using the same number of leaves, found with an exhaustive search. As those results were perfectly identical for every number of compared leaves, up to trees with 8 leaves and depth 4, we confirm that a genetic algorithm is appropriate for this problem.

Therefore, the resulting trees, presented in the next section, are validated close to the best available for the considered sets of \textit{tests} and \textit{labels}.

\section{Results and discussions} 
\label{res}

\subsection{Experimental setup}
We consider video sequences for evaluation from \gls{HEVC} and \gls{JEM} \glspl{CTC}~\cite{seregin_common_2014, suehring2016jvet}. The two video sets, described in Table~{\ref{table::seq_ctc}}, are identical except for class A  which is different and class F is optional in the \gls{JEM} \gls{CTC}. 
The test set video sequences are then encoded with both the \gls{HM} and \gls{JEM} encoders at four \gls{QP} $\in \{22, \, 27, \, 32, \, 37 \}$ in \gls{AI} coding configuration. The native \gls{IPM} signalling schemes in the \gls{HM} and \gls{JEM} represent the anchors while the modified with the our signalling schemes represent the proposed schemes. The performance of these latter is assessed in terms of average number of bits per \gls{IPM}, \gls{BD-BR}~\cite{bjontegaard2001calculation,bjontegaard2008improvements} and codec complexity with respect to the anchors.  

\subsection{Results and Analysis}

\subsubsection{Performance versus HEVC}

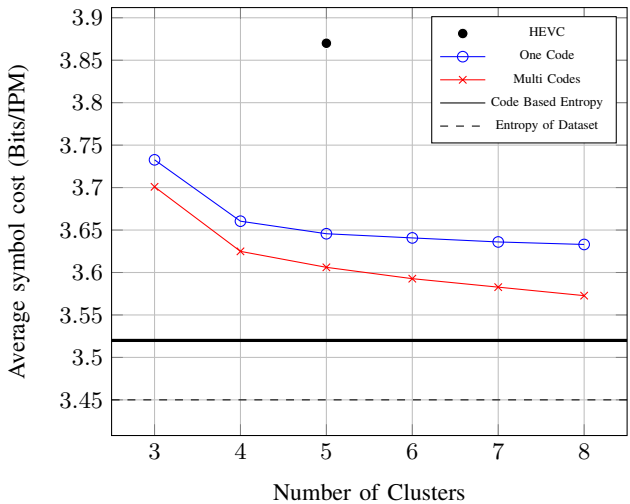
\begin{figure}[ht]
\centering
\begin{tikzpicture}
\begin{small}
\begin{axis}[%
scatter/classes={%
    c={mark=*,draw=white},
    a={mark=o,draw=blue},
    b={mark=x,draw=red},
    bbis={mark=.,draw=black},
    cbis={dashed,mark=.,draw=black},
    n={mark=-,draw=none}},
    xlabel={Number of Clusters},
    xtick={3, 4, 5, 6, 7, 8},
    ylabel={Average symbol cost (Bits/IPM)},
    ytick={3.45,3.5,3.55,3.6,3.65,3.7,3.75, 3.8, 3.85, 3.9},
    grid,
    legend style={font=\tiny}]
    
\addplot[scatter,draw=black,%
    scatter src=explicit symbolic]%
table[meta=label] {
x y label
5 3.87 c
    };
    \addlegendentry{HEVC}
    
\addplot[scatter,draw=blue,%
    scatter src=explicit symbolic]%
table[meta=label] {
x y label
3 3.7327 a
4 3.66042 a
5 3.64565 a
6 3.64073 a
7 3.63598 a
8 3.63301 a
    };
    \addlegendentry{One Code}
    \addplot[scatter,draw=red,%
    scatter src=explicit symbolic]%
table[meta=label] {
x y label
3 3.70091 b
4 3.6249 b
5 3.60609 b
6 3.59273 b
7 3.58275 b
8 3.57275 b
    };
    \addlegendentry{Multi Codes}
    \addlegendentry{Code Based Entropy}    
        \addplot[scatter,draw=white,%
    scatter src=explicit symbolic]%
table[meta=label] {
x y label
5 3.45 bbis
    };
    \addlegendentry{Entropy of Dataset}
    \draw[very thick] (axis cs:\pgfkeysvalueof{/pgfplots/xmin},3.52) -- (axis cs:\pgfkeysvalueof{/pgfplots/xmax},3.52);    
    \draw[dashed] (axis cs:\pgfkeysvalueof{/pgfplots/xmin},3.45) -- (axis cs:\pgfkeysvalueof{/pgfplots/xmax},3.45);       
\end{axis}
\end{small}
\end{tikzpicture}
\caption{Average bits/IPM for the two evaluated system setups for different number of clusters. The black lines represent the two limits described in the previous Section. }
\label{fig:bits_per_ipm_hevc}
\end{figure}

The first experiment consists in measuring the efficiency of the created signalling schemes on the testing datasets in bits/\gls{IPM}. This is performed outside the codec, simply based on the previously discussed dataset. Fig.~{\ref{fig:bits_per_ipm_hevc}} shows the performance in bits/\gls{IPM} of different signalling schemes created from the proposed list of \textit{tests}, \textit{labels} and \textit{codes} with different numbers of clusters from 3 to 8. The performance is provided for the proposed scheme in two configurations using one code and multiple codes for the different clusters. The proposed solution is compared to \gls{HEVC}, code-based entropy and entropy. We can notice that even with using less clusters, the proposed scheme enables better coding efficiency compared to \gls{HEVC}.

We can also notice that multiple codes configuration performs better than using only one code for all the clusters. Moreover, the multiple codes configuration converges faster toward the optimal solution (ie. the code-based entropy). The performance of the most efficient created system with 8 leaves is close by 0.05 bits/\gls{IPM} from the best achievable system. This confirms the efficiency of the considered \textit{tests} and \textit{labels}.

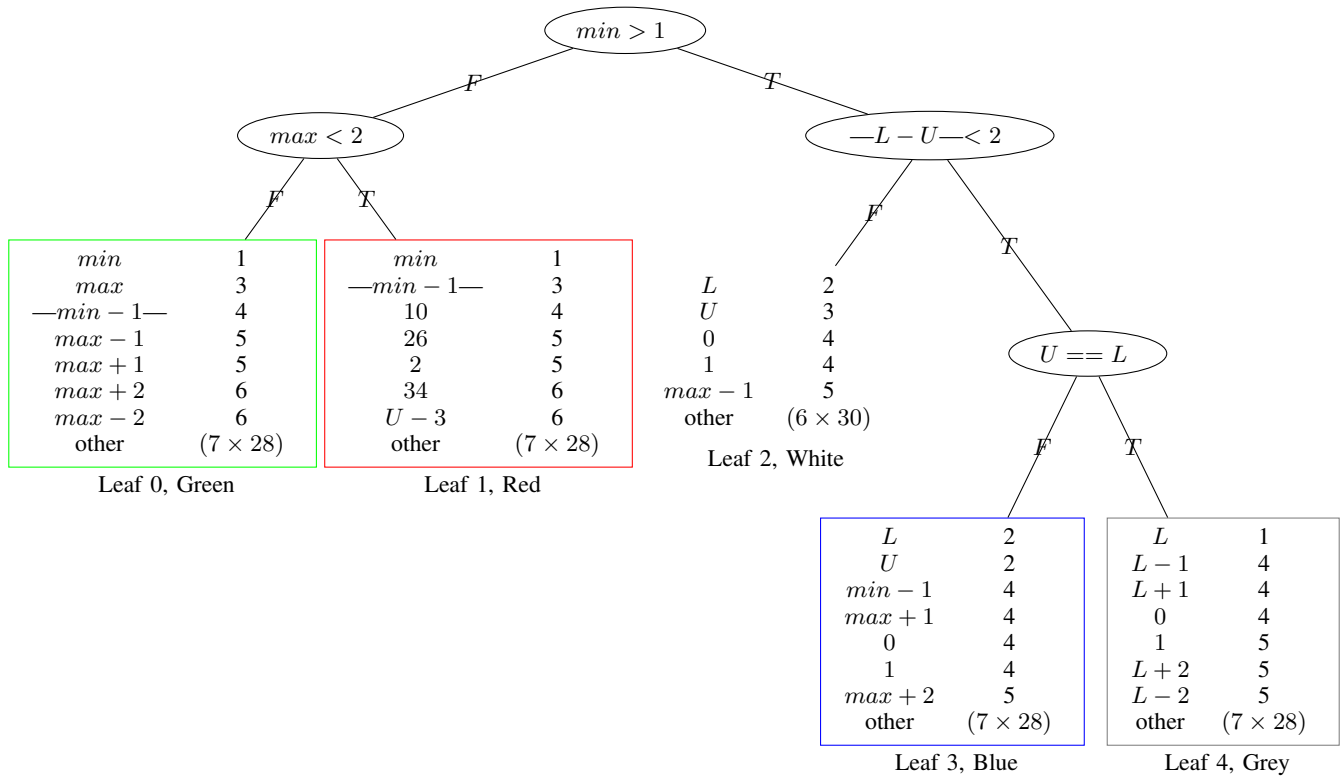
\begin{figure*}[ht]
\center
\tikzstyle{level 1}=[level distance=1.4cm, sibling distance=8.1cm]
\tikzstyle{level 2}=[level distance=2.9cm, sibling distance=4.2cm]
\tikzstyle{level 3}=[level distance=3.7cm, sibling distance=3.6cm]

\tikzstyle{bag} = [text centered, draw, ellipse, align=center]
\tikzstyle{end} = [minimum width=3pt, draw, align=center]

\begin{tikzpicture}[grow=down]
\small

\node[bag] { $min > 1$ }
child
{
	node[bag] { $max<2$ }
	child
	{
		node[end, draw=green, label=below: { Leaf 0, Green }] 
			{ 
			\begin{tabular}{c c}
			$min$ & 1\\
			$max$ & 3\\
			|$min-1$| & 4\\
			$max-1$ & 5\\
			$max+1$ & 5\\
			$max+2$ & 6\\
			$max-2$ & 6\\
			other & $(7\times 28)$\\
			\end{tabular}
			}
		edge from parent
			node[] { $F$ }
	}
	child
	{
		node[end, draw=red, label=below: { Leaf 1, Red }]
			{ 
			\begin{tabular}{c c}
				$min$ & 1\\
				|$min-1$| & 3\\
				$10$ & 4\\
				$26$ & 5\\
				$2$ & 5\\
				$34$ & 6\\
				$U-3$ & 6\\
				other & $(7\times 28)$
			\end{tabular}
			}
		edge from parent
			node[] { $T$ }
	}
	edge from parent
		node[] { $F$ }
}
child
{
	node[bag] { |$L-U$|$<2$ }
	child
	{
		node[end, draw=white, label=below: { Leaf 2, White }]
			{ 
			\begin{tabular}{c c}
				$L$ & 2\\
				$U$ & 3\\
				$0$ & 4\\
				$1$ & 4\\
				$max-1$ & 5\\
				other & $(6\times 30)$\\
			\end{tabular}
			}
		edge from parent
			node[] { $F$ }
	}
	child
	{
		node[bag] { $U==L$ }
		child
		{
			node[end, draw=blue, label=below: { Leaf 3, Blue }]
				{ 
				\begin{tabular}{c c}
					$L$ & 2\\
					$U$ & 2\\
					$min-1$ & 4\\
					$max+1$ & 4\\
					$0$ & 4\\
					$1$ & 4\\
					$max+2$ & 5\\
					other & $(7\times 28)$\\
				\end{tabular}
				}
			edge from parent
				node[] { $F$ }
		}
		child
		{
			node[end, draw=gray, label=below: { Leaf 4, Grey }]
				{ 
				\begin{tabular}{c c}
					$L$ & 1\\
					$L-1$ & 4\\
					$L+1$ & 4\\
					$0$ & 4\\
					$1$ & 5\\
					$L+2$ & 5\\
					$L-2$ & 5\\
					other & $(7\times 28)$\\
				\end{tabular}
				}
			edge from parent
				node[] { $T$ }
		}
		edge from parent
			node[] { $T$ }
	}
	edge from parent
		node[] { $T$ }
}
;
\end{tikzpicture}
\caption{Selected system for \gls{HEVC} shown as a decision tree. For each cluster the used code is shown below it. For clarity $min(U,L)$ and $max(U,L)$ are simply written $min$ and $max$. The colours refer to the clustering shown in Fig.~{\ref{fig:real_5leaf_clustering}}.} 
\label{fig:5leaf_decision_tree}
\end{figure*}

As \gls{HEVC} uses a signalling scheme with 5 clusters, the signalling scheme that we propose for integration in the \gls{HM} and comparison in terms of rate distortion is the one using 5 clusters. This scheme is presented as a binary decision tree in Fig.~{\ref{fig:5leaf_decision_tree}}. This figure gives the selected tests to perform clustering and the used codes by the five clusters. 

\begin{figure}[ht]
	\centering
	\subfloat[HEVC clustering]{\includegraphics[width= 0.5\linewidth]{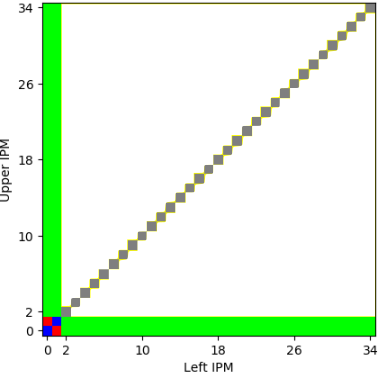}\label{fig:real_5leaf_clustering-hevc}} 
	\subfloat[Proposed clustering]{\includegraphics[width= 0.5\linewidth]{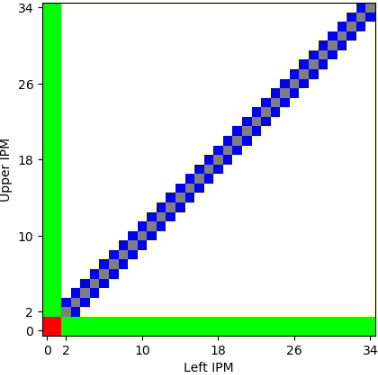}\label{fig:real_5leaf_clustering-pro}} 
	\caption{ The clustering of the 5-leaf signalling scheme. Each colour represents a cluster.}
	\label{fig:real_5leaf_clustering}
\end{figure}

The clustering representation of this scheme is illustrated in Fig.~{\ref{fig:real_5leaf_clustering-pro}} with respect to left and upper modes $ L$ and $U$. We can notice from this figure that the new labels and tests enable better clustering than \gls{HEVC} clustering illustrated in Fig.~{\ref{fig:real_5leaf_clustering-hevc}}.  Moreover, the clustering of the proposed scheme is close from the clustering derived with perfect {\it tests} illustrated in Fig.~\ref{fig:clustering-tests}, which asserts the efficiency of the proposed list of {\it tests}. Moreover, some \textit{tests} are used enabling the use of some specific \textit{labels}: for example the test $max < 2$ is used to leverage the numerical \textit{labels} 10, 26, 2 and 34 on the second leaf as they could not be used on the first leaf along with $max(L,U)$.

This signalling scheme uses 4 different \textit{codes} on the 5 different clusters where 3 of those 4 \textit{codes} use 7 \glspl{MPM}. The only cluster using a 5-\glspl{MPM} \textit{code} is the one where both \glspl{IPM} are angular and are not close to each other ($min(L,U)>1$ is true and $|L-U|<2$ is false, corresponding to the yellow cluster in Fig.~{\ref{fig:real_5leaf_clustering}}).
Another main difference with \gls{HEVC} is the use in the \glspl{MPM} list of the two non-angular modes, either through the use of the \textit{labels} $min(L,U)$ and $|min(L,U)-1|$ on the leaves 0 and 1 (as, on these leaves, per the first test we know that the minimum between $L$ and $U$ is either 0 or 1, $min(L,U)$ and $|min(L,U)-1|$ refer to 0 and 1 or 1 and 0) or explicitly as 0 and 1 on the other leaves.

%

This system has been selected to be tested in the HM codec under the \gls{CTC} in \gls{AI} coding configuration. Table~{\ref{table::res_rate_ctc}} gives the \gls{BD-BR} of the proposed codes with respect to the two anchors \gls{HM} and \gls{JEM}, respectively.  For \gls{HEVC}, the proposed coding scheme enables -0.18\% bitrate gain on average on all test sequences. Moreover, all sequences benefit from this new signalling scheme. Such gains are extremely interesting as they do not add significant complexity at both encoder and decoder. The encoding and decoding run times are presented in Table~{\ref{tab:enc_dec_times_ctc}}.
The results published in~\cite{reuze2016intra} and \cite{reuze2017cluster} emphasis better \gls{BD-BR} results ($-0.4\%$) with similar systems, it has to be noted that their implementations do not consider the complexity introduced by the increased number of \gls{MPM} for the \gls{RD} search in \gls{HEVC}. The implementation for the system proposed in this work asserts that the number of modes tested in \gls{RD} remains exactly the same as for the anchor.  This coding efficiency can be further improved by considering coding schemes with higher number of clusters up to 8 clusters at the expense of slight increase in the codec complexity.

\subsubsection{Performance versus JEM}

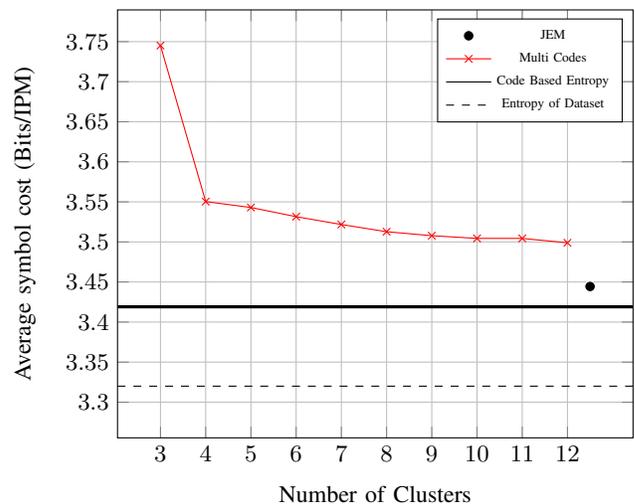
\begin{figure}[ht]
\centering
\begin{tikzpicture}
\begin{small}
\begin{axis}[%
scatter/classes={%
    c={mark=*,draw=white},
    b={mark=x,draw=red},
    bbis={mark=.,draw=black},
    cbis={dashed,mark=.,draw=black},
    n={mark=-,draw=none}},
    xlabel={Number of Clusters},
    xtick={3, 4, 5, 6, 7, 8, 9, 10, 11, 12}, 
    ylabel={Average symbol cost (Bits/IPM)},
    ytick={3.3,3.35,3.4,3.45,3.5,3.55,3.6,3.65,3.7,3.75, 3.8, 3.85, 3.9},
    grid,
    legend style={font=\tiny}]
    
\addplot[scatter,draw=black,%
    scatter src=explicit symbolic]%
table[meta=label] {
x y label
12.5 3.4444 c
    };
    \addlegendentry{JEM}   

    \addplot[scatter,draw=red,%
    scatter src=explicit symbolic]%
table[meta=label] {
x y label
3	3.74509	b
4	3.55034	b
5	3.54299	b
6	3.53154	b
7	3.52178	b
8	3.51277	b
9	3.50779	b
10	3.50442	b
11	3.50442	b
12	3.49892	b
    };
    \addlegendentry{Multi Codes}
    \addlegendentry{Code Based Entropy}
    
        \addplot[scatter,draw=white,%
    scatter src=explicit symbolic]%
table[meta=label] {
x y label
5 3.3 bbis
    };
    \addlegendentry{Entropy of Dataset}
    \draw[very thick] (axis cs:\pgfkeysvalueof{/pgfplots/xmin},3.41905) -- (axis cs:\pgfkeysvalueof{/pgfplots/xmax},3.41905);    
    \draw[dashed] (axis cs:\pgfkeysvalueof{/pgfplots/xmin},3.32) -- (axis cs:\pgfkeysvalueof{/pgfplots/xmax},3.32);    
\end{axis}
\end{small}
\end{tikzpicture}
\caption{Average bits/IPM for the two evaluated system setups for different number of clusters. The black lines represent the two limits described in the previous Section. The \gls{JEM} signalling scheme is put here for comparison but actually uses 460 clusters. }
\label{fig:bits_per_ipm_jem}
\end{figure}

Fig.~{\ref{fig:bits_per_ipm_jem}} gives the performance of the proposed signalling scheme in average number of bits/\gls{IPM} for different number of clusters in comparison with the code-based entropy and entropy. The performance of the proposed signaling scheme is close to the code-based entropy with using only 3 contexts ($L$, $U$ and $UL$) where \gls{JEM} uses 5 contextual information since it also uses $BL$ and $UR$.
Moreover, the \gls{JEM} uses 460 different leaves and 196 different labellings while our signalling scheme relies on only  12 leaves. 

An interesting observation that cannot be seen simply by looking at the number of bits/\gls{IPM} in Fig.~{\ref{fig:bits_per_ipm_jem}} is that the trees only use \textit{tests} and \textit{labels} relying on $L$ and $U$ up to 7 clusters (and therefore $UL$ context is not used in trees with less than 8 clusters). It means that the \textit{tests} based on $UL$ provide much less information than those based on $L$ and $U$, and this explains why so many clusters are required by the \gls{JEM} to efficiently use $BL$ and $UR$.

The scheme with 12-leaf tree is only 0.05 bits/\gls{IPM} away from the \gls{JEM} performance, we think appropriate to test it under the \glspl{CTC} and see if our simplification of the \gls{JEM} signaling scheme provides good \gls{RD} performance. As the \gls{JEM} signalling scheme uses 4 different \gls{CABAC} contexts, our proposed signalling scheme also relies  on only  4 \gls{CABAC} contexts. On each leaf one \gls{CABAC} is used on the first bit of the code. The leaves are grouped so that leaves with similar probabilities for this first bit uses the same \gls{CABAC} context.

The 12-leaf tree enables similar coding efficiency as the \gls{JEM} under the \glspl{CTC} in \gls{AI} coding configuration with +0.03\% \gls{BD-BR} increase on average on the test sequences, as illustrated in Table~{\ref{table::res_rate_ctc}}.

The encoding and decoding times are presented in Table~\ref{tab:enc_dec_times_ctc}. The new code used for \gls{HEVC} slightly increases the codec complexity caused by the 3 used \gls{CABAC} contexts while only one context is used by \gls{HEVC}. For the \gls{JEM}, the complexity remains similar with the anchor at both encoder and decoder since they use the same number of 5 \gls{CABAC} contexts. 
This complexity performance proves the method's efficiency to provide efficient and simple signalling schemes. Moreover, the proposed codes are also simpler than the one used in the \gls{JEM} as they only use a list of \glspl{MPM} followed by a unique code-length, while the \gls{JEM} code uses 3 different code-lengths for its non-predicted modes.

\begin{table}[t]
\renewcommand{\arraystretch}{1.1}
\centering
\caption{BD-rate gains in \gls{AI} coding configuration under \gls{CTC}}
\label{table::res_rate_ctc}
\begin{tabular}{| l | l c c |}
\hline
Class & Name & \gls{HEVC} & \gls{JEM} \\  [0.5ex]
\hline
\hline
\multirow{5}{*}{A1} & \textit{CampfireParty} & - & 0.07\%  \\  [0.5ex]
 & \textit{Drums} & - & -0.08\% \\  [0.5ex]
 & \textit{Tango} & - & -0.04\%  \\  [0.5ex]
 & \textit{ToddlerFountain} & - & 0.13\% \\  [0.5ex]
 \hline
 \multirow{5}{*}{A2} & \textit{CatRobot} & - & -0.02\% \\  [0.5ex]
 & \textit{DaylightRoad} & - & 0.09\% \\  [0.5ex]
 & \textit{RollerCoaster} & - & -0.05\% \\  [0.5ex]
 & \textit{TrafficFlow} & - & 0.03\% \\  [0.5ex]
 \hline
\multirow{5}{*}{A} & \textit{NebutaFestival} & -0.07\% & - \\  [0.5ex]
 & \textit{PeopleOnStreet} & -0.27\% & -\\  [0.5ex]
 & \textit{SteamLocomotiveTrain} & -0.12\% & - \\  [0.5ex]
 & \textit{Traffic} & -0.15\% & - \\  [0.5ex]
 \hline
\multirow{5}{*}{B} & \textit{BasketballDrive1} & -0.19\% & 0.07\% \\  [0.5ex]
 & \textit{BQTerrace} & -0.13\% & 0.21\% \\  [0.5ex]
 & \textit{Cactus} & -0.18\% & 0.01\% \\  [0.5ex]
 & \textit{Kimono1} & -0.17\% & -0.01\% \\  [0.5ex]
 & \textit{ParkScene} & -0.14\% & 0.14\% \\  [0.5ex]
 \hline
\multirow{4}{*}{C} & \textit{BasketballDrill} & -0.27\% & 0.18\% \\  [0.5ex]
& \textit{BQMall} & -0.12\% & 0.02\% \\  [0.5ex]
& \textit{PartyScene} & -0.09\%  & 0.10\% \\  [0.5ex]
& \textit{RaceHorses} & -0.18\% & -0.04\% \\  [0.5ex]
\hline
\multirow{4}{*}{D} & \textit{BasketballPass} & -0.06\%  & 0.06\% \\  [0.5ex]
 & \textit{BlowingBubbles} & -0.09\%  & 0.00\% \\  [0.5ex]
 & \textit{BQSquare} & -0.09\% & 0.16\% \\  [0.5ex]
 & \textit{RaceHorses} & -0.18\% & 0.02\% \\  [0.5ex]
 \hline
\multirow{3}{*}{E} & \textit{FourPeople} & -0.24\%  & -0.05\% \\  [0.5ex]
& \textit{Johnny} & -0.25\% & -0.13\% \\  [0.5ex]
& \textit{KristenAndSara} & -0.29\% & -0.15\% \\  [0.5ex]
\hline
\multirow{5}{*}{F} & \textit{BasketballDrillText} & -0.20\%  & - \\  [0.5ex]
 & \textit{ChinaSpeed} & -0.27\% & - \\  [0.5ex]
 & \textit{SlideEditing} & -0.14\%  & - \\  [0.5ex]
 & \textit{SlideShow} & -0.41\% & - \\  [0.5ex]
\hline
- & \textbf{Average} & \textbf{-0.18\%}  & \textbf{+0.03\%} \\  [0.5ex]
\hline
\end{tabular}
\end{table}


\section{Dynamic lists} 
\label{sec:dynamiclist}
We can notice from the previous section that the coding performance of the proposed solution is quite similar to the anchor \gls{JEM}. Moreover, the BD-BR increase for some sequences like  {\it BQTerrace} or {\it BasketballDrill} is mainly caused to the distribution of \glspl{IPM} in these sequences, which is not well leveraged by the proposed coding scheme. In this section we propose a combination of dynamic lists and decision tree to build a more efficient coding scheme of the  \gls{JEM} \glspl{IPM}.

\subsection{Labels Used in Dynamic Lists} 
The \gls{JEM} uses 21 labels in its dynamic list: the 5 neighbouring \gls{IPM}s $L$, $U$, $UL$, $UR$ and $BL$, the corresponding directional neighbours of those modes $L\pm 1$, $U\pm 1$, ... and 6 numerical labels, 0, 1, 2, 18, 34 and 50. To create a more efficient dynamic list than the one defined in the \gls{JEM}, it is necessary to include new labels that can enhance the prediction. 

Labels are selected in the same way as in the previous section: the existing labels are extended in order to be able to capture more information. The labels presented in our proposed list consist of the \gls{JEM} modes and more directional neighbours for the present modes ($L\pm 3$, $L\pm 4$, ...). All directional modes are also used, giving a final list of 117 labels, including the 67 numerical labels and 50 labels based on contextual information.

To perfectly reproduce the performance of the \gls{JEM} code, labels that can replicate the preferred list are needed. Those would be labels based on the numerical values of the previously used modes in the \gls{MPM} list. However, using those labels increases the search for the best order of labels by a factor of 10. Due to this increase in complexity it was chosen not to use such modes in the presented dynamic lists.
\subsection{Creating Dynamic Lists}
A dynamic list is an ordering of a list of labels. From the list of 117 available labels there are therefore $117!=4\times 10^{192}$ possible dynamic lists. To handle such levels of complexity, a genetic algorithm is used to order the labels. The genetic algorithm works as a guided random search. One candidate solution is an order of labels. Its cost is evaluated by creating the probability histogram of indexes containing 67 values. On this histogram all available codes are tested, and the most efficient  one is kept. The mutations for this genetic algorithm consist in operations changing the order of labels, such as swapping two labels, inserting a label in a random position, etc. As there is still no certainty on the optimality of the genetic algorithm, safety checks are performed. This latter performs changes on the final lists obtained to see if its performance can be enhanced. The overall order is kept but the cost is measured for every change of position for individual {\it label} in the list.
\begin{table}[t]
\renewcommand{\arraystretch}{1.1}
\centering
\caption{encoding and decoding times in \gls{AI} coding configuration in \gls{CTC} compared to respective anchors.}
\label{tab:enc_dec_times_ctc}
\begin{tabular}{| l | c c c c |}

\hline
&  \multicolumn{2}{c}{\gls{HEVC}} & \multicolumn{2}{c|}{\gls{JEM}}\\ 
Class & Enc. Time & Dec. Time & Enc. Time & Dec. Time \\
\hline
\hline
 A1 & - & - & 100\% & 99\% \\ 
 A2 & - & - & 101\% & 101\% \\ 
 A & 103\% & 106\% & - & - \\ 
B & 104\% & 103\% & 100\% & 100\% \\  
C & 103\% & 98\% & 100\% & 100\% \\  
D & 104\% & 101\% & 100\% & 100\% \\  
E & 104\% & 99\% & 101\% & 100\% \\  
F & 104\% & 101\% & - & - \\ 
\hline
\textbf{Average} & \textbf{104\%}  & \textbf{102\%} & \textbf{100\%}  & \textbf{100\%} \\ 
\hline
\end{tabular}
\end{table}

\subsection{Combining Dynamic Lists and Decision Trees}

One of the most advantages presented in our previous section is the use of multiple codes in the designed signalling schemes. Using different codes allows better fit of the probability distributions. Therefore, we propose to combine a decision tree with dynamic lists to enable using different codes and different predictions. On each leaf of the decision tree, one dynamic list can be created with its corresponding code. This way both the prediction and the code better fit the different clusters.

To create those decision trees, the proposed tests are based on the same contextual information as the labels. It appeared that having tests solely based on $L$ and $U$ was sufficient for creating efficient decision trees, as those two labels alone captured most of the information. Moreover, including additional neighbour modes would require more complex decision tree to leverage this contextual information. In our approach, one dynamic list is created for each leaf, and since this is a time consuming process, as explained in the previous section, small decision trees are preferred. 

\subsection{Experimental results}
\begin{figure*}[htbp]
\tikzstyle{level 1}=[level distance=1.6cm, sibling distance=9.3cm]
\tikzstyle{level 2}=[level distance=6cm, sibling distance=4.8cm]

\definecolor{salmon}{RGB}{255, 160, 122}

\tikzstyle{bag} = [text centered, draw, ellipse, align=center]
\tikzstyle{end} = [minimum width=3pt, draw, ultra thick, align=center]

\begin{tikzpicture}[grow=down]
\node[bag] { $min(L,U)>1$ }
child
{
	node[bag] { $max(L,U)<2$ }
	child
	{
	node[end, draw=green, label=below: { \begin{tabular}{c}
 Leaf 0, $P(b_0)=58\%$ \end{tabular} } ] %
				{\begin{tabular}{c c} 
2	&	\textbf{0}0	\\
2	&	\textbf{0}1	\\
\hline
3	&	\textbf{1}00	\\
\hline
	&	\textbf{1}0100	\\
5( $\times$ 3)	&	\textbf{1}0101	\\
	&	\textbf{1}0110	\\
\hline
	&	\textbf{1}~011100	\\
$7(\times 11)$	&	~~~\vdots	\\
	&	\textbf{1}~100110	\\
\hline
	&	\textbf{1}1~001110	\\
$8(\times 50)$	&	~~~~\vdots	\\
	&	\textbf{1}1~111111	\\
\end{tabular}}
			edge from parent
				node { \textit{F} }
	}
	child
	{
		node[end,draw=red, label=below: { \begin{tabular}{c}
 Leaf 1, $ P(b_0)=77\%$  \end{tabular} } ]%
				{\begin{tabular}{c c} 
1	&	0	\\
\hline
\hline
3	&	1\textbf{0}0	\\
5	&	1\textbf{0}100	\\
5	&	1\textbf{0}101	\\ 
\hline
	&	1\textbf{0}11~000	\\
  7( $\times$ 8) & ~~~~~~~~~\vdots \\
	&	1\textbf{0}11~111	\\
	\hline
	&	1\textbf{1}0~0000	\\
$7(\times 16)$	&	~~~~~~~\vdots	\\
	&	1\textbf{1}0~1111	\\
\hline 
	&	1\textbf{1}1~00000	\\
$8(\times 25)$	&	~~~~~~~\vdots	\\
	&	1\textbf{1}1~11000	\\
\hline 
	&	1\textbf{1}111~0010	\\
$9(\times 14)$	&	~~~~~~~~~~\vdots	\\
	&	1\textbf{1}111~1111	\\
\end{tabular}}
		edge from parent
			node { $T$ }
	}
	edge from parent
		node { $F$ }
}
child
{
		node[bag] { |$L-U$|$\%63<3$ } 
		child
		{
				node[end, draw=yellow, xshift=-0.3cm, label=below: { \begin{tabular}{c}
 Leaf 2, $p(b_0)=64\% $ \end{tabular} } ]
					{\begin{tabular}{l c} 
2	&	\textbf{0}0	\\
3	&	\textbf{0}10	\\
3	&	\textbf{0}11	\\
\hline 
\hline 
\multirow{2}{*}{  4( $\times$ 2)  }		&	\textbf{1}000	\\
	&	\textbf{1}001	\\
	\hline
	&	\textbf{1}01~000	\\
$6(\times 8)$	&	~~~~~~~\vdots	\\
	&	\textbf{1}01~111	\\
\hline 
	&	\textbf{1}10~0000	\\
$7(\times 10)$	&	~~~~~~~\vdots	\\
	&	\textbf{1}10~1001	\\
\hline			 
	&	\textbf{1}1~010100	\\
$8(\times 44)$	&	~~~~~~\vdots	\\
	&	\textbf{1}1~111111	\\
 \end{tabular}}
				edge from parent
					node { $F$ }
	}
	child
	{
			node[end, draw=blue, xshift=-0.5cm, label=below: { \begin{tabular}{c}
 Leaf 3,  $P(b_0)=76\%$ \end{tabular} } ] %
				{\begin{tabular}{l c} 
1	&	0	\\
\hline			
\hline
3	&	1\textbf{0}0	\\
4	&	1\textbf{0}10	\\
5	&	1\textbf{0}110	\\
6	&	1\textbf{0}1110	\\
6	&	1\textbf{0}1111	\\
\hline			
\hline 
	&	1\textbf{1}0~000	\\
{ 6( $\times$ 6) }	&	~~~~~~~\vdots	\\
	&	1\textbf{1}0~101	\\
	\hline
	&	1\textbf{1}~01100	\\
$7(\times 6)$	&	~~~~~~\vdots	\\
	&	1\textbf{1}~10001	\\
\hline			 
	&	1\textbf{1}10~0100	\\
$8(\times 7)$	&	~~~~~~~~~\vdots	\\
	&	1\textbf{1}10~1010	\\
\hline 
	&	1\textbf{1}1~010110	\\
$9(\times 42)$	&	~~~~~~~~\vdots	\\
	&	1\textbf{1}1~111111	\\ 
 \end{tabular}}
		edge from parent
			node { $T$ }
	}
	edge from parent
		node { $T$ }
}
;
\end{tikzpicture}
\caption[Number of bits used on each leaf of the decision tree for dynamic lists.]{Number of bits used on each leaf of the decision tree for dynamic lists. On the left is the number of bits, multiplied by the number of symbols when necessary. Bits in bold are coded using a \gls{CABAC} context. One \gls{CABAC} context is used by leaf.}
\label{fig:codes_4leaf_dyna}
\end{figure*}
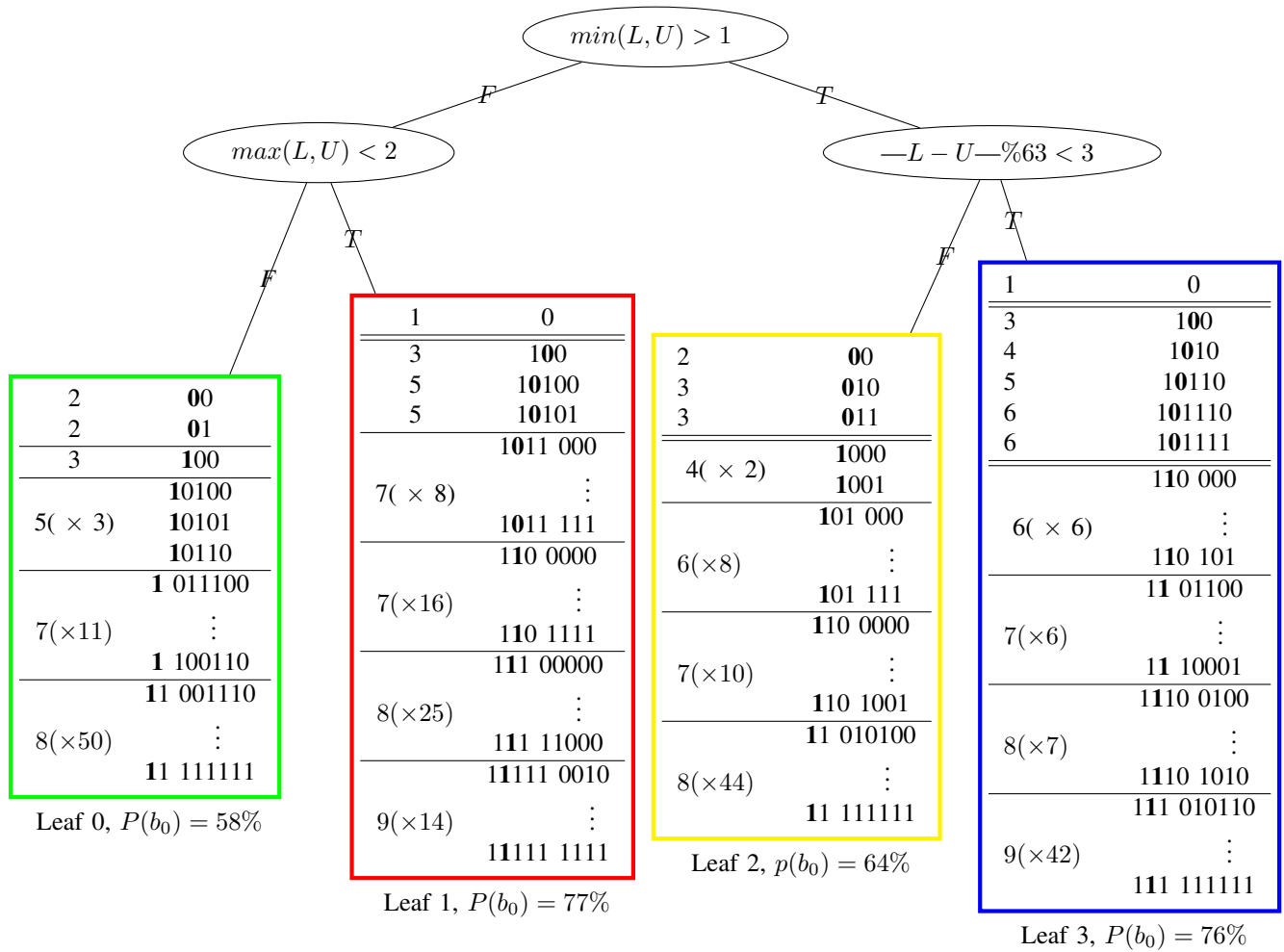

The designed trees incorporate 4 leaves, each linked to a dynamic list, as illustrated in Fig.~\ref{fig:codes_4leaf_dyna}. Each dynamic list (leaf) employs a distinct code. Additionally, a single \gls{CABAC} context is employed for the first bit of each code. Consequently, the designed schemes utilize a total of 4 distinct \gls{CABAC} contexts, which is fewer than the five contexts employed by the \gls{JEM}. The training process for constructing these dynamic lists is an iterative one, involving multiple training passes. In the initial pass, samples from the \gls{JEM} native codes are used, while in the subsequent passes, samples encoded by the codes generated in the preceding pass are utilized. Furthermore, it's important to note that different datasets are employed in the training process for constructing the dynamic lists in each pass.
\begin{table}[htbp]
\renewcommand{\arraystretch}{1.1}
	\caption{Results in bits/\gls{IPM} of the different pass for dynamic lists using a decision tree. The reference (Ref) in Pass 1 is the \gls{JEM}. The reference of each row then corresponds to the New of the previous row. Delta is $\frac{New - Ref}{Ref}$.}
	\label{tab:dynalist_bitsIPM}
		\centering 
	\begin{tabular}{| c | c c c c |}
	
	\hline
	&	JEM	&	Ref	&	New	&	Delta	\\
	\hline
	\hline
Pass 1	&	3.416	&	3.416	&	3.436	&	0.59\%	\\
Pass 2	&	3.426	&	3.287	&	3.251	&	-1.10\%	\\
Pass 3	&	3.414	&	3.299	&	3.283	&	-0.48\%	\\
Pass 4	&	3.410	&	3.287	&	3.282	&	-0.15\%	\\
\hline
	\end{tabular}
\end{table}
Table~\ref{tab:dynalist_bitsIPM} shows the results in bits/IPM of the proposed scheme after each pass. The first pass corresponds to data encoded with the \gls{JEM} encoder. On this set, the designed tree is not as efficient as the \gls{JEM} in terms of bits/IPM. Training sequences are encoded using this designed decision tree, to provide a training set for the second pass. A second dynamic tree is designed from this list. It uses the same tests as the previous one with different order of modes for the \gls{MPM} lists and new codes at each leaf. The dynamic lists built from the second pass enable a reduction of 1.1\% in terms of bits/IPM. For the third pass, a gain in bits/IPM is still observed but it is more than halved compared with the previous pass. As the tree designed for a fourth pass shows an even smaller improvement in bits/IPM. At this stage, convergence seems reached, and we decided to not to perform further passes, nor to implement this tree.  Fig.~\ref{fig:codes_4leaf_dyna} shows the tree trained in the third pass as well as the codes (up to the first fixed length code) used at each leaf. This tree achieved 3.283 bits/IPM for the training set resulting from the pass 3, and 3.287 on the training set of the 4th pass.

Table \ref{tab:dyna_tree_bdrate} presents the results in BD-rate of the successively implemented trees under the CTCs. The first observation is that, even if the tree designed from the first pass performs theoretically worse than the JEM, it achieves a BD-rate gain of 0.04\%. This is a small gain, but considering that it uses one less CABAC context than the JEM it is promising. The use of multiple passes is then proved to be efficient, since the gains are increased to 0.09\% for the third pass. The small decrease in BD-rate between the second and third pass confirms that convergence has been reached and that no further pass should be able to reduce the coding efficiency.

\begin{table}[h]                         
\centering                         
\renewcommand{\arraystretch}{1.1}
\caption[\gls{BD-BR} results of dynamic tree against \gls{JEM}]{Results in \gls{BD-BR} of the tree using dynamic labelling on each leaf, against \gls{JEM} in \gls{AI}. The tree designed in the first pass uses data from \gls{JEM}7 encoding, and then each tree uses data encoded using the previous pass.}
\label{tab:dyna_tree_bdrate}                 
\begin{tabular}{| l | c c c |}                      
\hline                           
 Class & Pass 1 & Pass 2 & Pass 3 \\ [0.5ex]            
\hline                       
\hline                         
 \textit{A1} & -0.02\% & -0.06\% & -0.07\% \\ [0.5ex]               
  \textit{A2} & 0.00\% & -0.07\% & -0.07\% \\ [0.5ex]               
  \textit{B} & 0.7\% & 0.03\% & 0.02\% \\ [0.5ex]               
  \textit{C} & -0.05\% & -0.06\% & -0.15\% \\ [0.5ex]                                       
 \textit{D} & -0.06\% & -0.11\% & -0.10\% \\ [0.5ex]               
  \textit{E} & -0.24\% & -0.29\% & -0.23\% \\ [0.5ex]               
  \textit{F} & 0.00\% & -0.14\% & -0.11\% \\ [0.5ex]                
\hline                         
 \textbf{Average} & \textbf{-0.04\%} & \textbf{-0.08\%} & \textbf{-0.09\%} \\ [0.5ex]               
\hline
\end{tabular}                       
\end{table}         

With a particular list, some optimisations can be made to reduce the number of tests made to create the dynamic list. These optimisations already performed by the JEM. Such optimisations are not considered for the presented lists, which is why no consideration should be taken for the complexity. An example of optimisation is, to stop using the dynamic list as soon as the remaining modes in the list are all coded using the same number of bits. This has no impact on the performance but reduces the number of tests performed and therefore the complexity. The same kind of optimisations can be made on all leaves. After these optimisations, the designed system should have the same complexity as the JEM, as it has been made sure that no extra modes are tested during the RD-search. Compared to recent methods proposed to encode the \glspl{IPM} \cite{8351605, 9115218}, our technique enables coding performance in the same range of 0.1\% \gls{BD-BR} gain under the JEM codec without increasing the complexity at both encoder and decoder sides. Under the HM software, the \gls{BD-BR} gain of 0.18\% is even higher with a slight complexity increase of 104\% and 102\% at encoder and decoder sides, respectively. These results clearly show the ability of the proposed method to derive efficient predictive coding schemes to encode \glspl{IPM}. 

\section{Conclusion}
\label{concl}

In this paper, we have developed a discrete prediction coding model that leverages contextual information in three key steps: prediction, clustering, and coding. To enhance these steps, we have introduced novel \textit{labels} for prediction, \textit{tests} for clustering, and \textit{codes} for coding. We employ a genetic algorithm to select the optimal combinations of \textit{labels}, \textit{tests}, and \textit{codes} that minimize the rate cost. This modeling approach has been applied to derive optimal coding schemes for \glspl{IPM} in both the \gls{HM} and \gls{JEM} reference software. It allows approaching the code-based entropy while increasing the number of clusters. These newly proposed coding schemes have been seamlessly integrated into the \gls{HM} and \gls{JEM} codecs. Under the \gls{HEVC} \glspl{CTC}, our proposed scheme with 5 clusters has yielded a coding efficiency improvement of -0.18\% while maintaining the same complexity level. In the case of the \gls{JEM}, the new scheme featuring 12 clusters has achieved coding performance comparable to the anchor scheme. Additionally, our research has demonstrated that combining dynamic lists with a simple decision tree results in superior coding performance compared to the \gls{JEM} under the same complexity constraints.

\def\url#1{}
\bibliographystyle{IEEEtran}
\bibliography{merged-biblio.bib}

\end{document}